\documentclass[AMA,STIX1COL]{WileyNJD-v2}
\usepackage{moreverb}
\usepackage{amssymb, amsmath}
\usepackage{graphicx}
\usepackage{subcaption}
\usepackage{bm}
\usepackage{verbatim}
\usepackage{caption}
\usepackage{subcaption}

\usepackage{tikz, tikzscale}
\usepackage{placeins}

\newcommand\BibTeX{{\rmfamily B\kern-.05em \textsc{i\kern-.025em b}\kern-.08em
T\kern-.1667em\lower.7ex\hbox{E}\kern-.125emX}}

\articletype{Article Type}%

\received{<day> <Month>, <year>}
\revised{<day> <Month>, <year>}
\accepted{<day> <Month>, <year>}

\begin{document}

\title{Per-event Uncertainty Quantification for Flow Cytometry using Calibration Beads}
\author[1,2]{Prajakta Bedekar*}
\author[3]{Megan A Catterton}
\author[3]{Matthew DiSalvo}
\author[3]{Gregory A Cooksey}
\author[1]{Anthony J Kearsley}
\author[1]{Paul N Patrone}

\authormark{Bedekar \textsc{et al}}

\address[1]{\orgdiv{Applied and Computational Mathematics Division, Information Technology Laboratory}, \orgname{National Institute of Standards and Technology}, \orgaddress{\state{Gaithersburg, Maryland}, \country{USA}}}

\address[2]{\orgdiv{Department of Applied Mathematics and Statistics}, \orgname{Johns Hopkins University}, \orgaddress{\state{Baltimore, Maryland}, \country{USA}}}

\address[3]{\orgdiv{Microsystems and Nanotechnology Division, Physical Measurement Laboratory}, \orgname{National Institute of Standards and Technology}, \orgaddress{\state{Gaithersburg, Maryland}, \country{USA}}}

\corres{*Prajakta Bedekar. \email{pbedeka1@jh.edu}}

\abstract[Abstract]{Flow cytometry measurements are widely used in diagnostics and medical decision making. Incomplete understanding of sources of measurement uncertainty can make it difficult to distinguish autofluorescence and background sources from signals of interest. Moreover, established methods for modeling uncertainty overlook the fact that the apparent distribution of measurements is a convolution of the inherent the population variability (e.g., associated with calibration beads or cells) and instrument induced-effects.  Such issues make it difficult, for example, to identify signals from small objects such as extracellular vesicles. To overcome such limitations, we formulate an explicit probabilistic measurement model that accounts for volume and labeling variation, background signals and fluorescence shot noise. Using raw data from routine per-event calibration measurements, we use this model to separate the aforementioned sources of uncertainty and demonstrate how such information can be used to facilitate decision-making and instrument characterization.}


\jnlcitation{\cname{%
\author{Bedekar}, 
\author{Catterton}, 
\author{DiSalvo}, 
\author{Cooksey},
\author{Kearsley}, and 
\author{Patrone}} (\cyear{20--}), 
\ctitle{Title}, \cjournal{journal}, \cvol{volume}.}

\maketitle

\footnotetext{\textbf{Abbreviations:} MESF, Molecules of Equivalent Soluble Fluorochrome; MFI, Mean Fluorescence Intensity}

%\twocolumn

\section{Introduction}

\renewcommand{\thefootnote}{\roman{footnote}}

Flow cytometry is a widely used technique in biomedical research and diagnostics \cite{mizrahi2018quantitative}. A complete understanding and model of uncertainty in cytometers remains an open question. In particular, cytometers convolve inherent population variability with instrument-induced uncertainties, but to the best of our knowledge, there exists no comprehensive analysis that accounts for the interplay \textit{between} these effects.  This has made it difficult, for example, to determine when a dim signal is due to background events or beads.  Similarly, for dim objects it is challenging to quantify what fraction of an observed signal is likely due to autofluorescence.  Fundamentally these problems arise from an inability to realize uncertainty decomposition or deconvolution, wherein the effects of the instrument are separated from those of the population.

In an effort to overcome such problems, previous works have developed models that account for instrument-specific effects,
e.g. due to background \cite{steen1992noise,wang2017flow,wang2017standardization}. Importantly, these models lead to data analysis strategies that are straightforward to implement using widely available reference materials such as calibration beads \cite{mittag2009basics,perfetto2012quality,finck2013normalization,byrd2015polystyrene}. However, such methods account for few sources of uncertainty and still do not fully address the issue of uncertainty propagation across the measurement process. Thus, there is still a need for tools that overcome this problem.

To address this need, we construct and analyze a probabilistic, per-event model of the measurement process that builds upon previous works by explicitly accounting for inherent variation in the measurand volume, randomness in labeling concentration, and autofluorescence, in addition to more routinely studied  optical effects and background signals\cite{overton1988modified, hulspas2009considerations}.    Given this model, we then formulate an optimization routine that \textit{isolates and quantifies}  the relative impact of these effects using the raw, per-event data associated with a calibration standard.   We validate this analysis using synthetic data and experimental measurements from a commercial flow cytometer.

A counterintuitive and key finding of this work is that autofluorescence facilitates and is even fundamental to uncertainty decomposition and modeling in cytometry.  Physically, this is justified by noting that autofluorescence is an inherent property of a material (e.g.\ bead, or ideally, a cell), that scales with the \textit{amount} of material, i.e.\ the volume.  In contrast, labeling entails an additional chemical reaction that deposits material randomly, which introduces additional uncertainty in the measurement process.  Thus, it stands to reason that measurements of unlabeled, autofluorescent particles allow one to distinguish volume variation from stochasticity in biomarker concentration, a fact that we demonstrate mathematically.  In later sections, we also discuss how this perspective, which is at odds with conventional wisdom, points to novel and useful experiments that leverage autofluorescence to extract more information from cytometry measurements than was previously believed possible.% \textcolor{red}{put references here}.

A related theme of this work is the observation that \textit{per-event} modeling is a more powerful and direct approach to characterizing cytometers than methods based on statistical moments (i.e.\ means and variances) of populations \cite{hoffman2007characterization,chase1998resolution}.   In particular, we show that the latter can be derived from general properties of our per-event formulation.  This explains the general success of existing models for analyzing cytometry calibration data, but by the same token, it points to their shortcomings.  For example, without reference to a more granular per-event model, existing moment-based modeling generally lacks the ability to distinguish population and instrument based effects.  Similarly, we show that the latter permits per-event uncertainty estimates of fluorophore concentrations, a task that, to the best of our knowledge, has yet to be fully addressed.

The rest of this manuscript is organized as follows.  Section \ref{Section:Theory} gives an overview of our main theoretical ideas and provides details on the experiments to which our data analysis is applied.  Section \ref{Section:Results} validates our analysis in the context of both synthetic and experimental detail, while Section \ref{Section:Discussion} discusses these results in the context of past works and possible extensions.  Appendices provide additional details on theoretical aspects of this work.

\section{Theory and Methods}
\label{Section:Theory}
Instrument effects arising from the measurement process, as well as population variability in the measurand,
determine how the per-event MESF maps to fluorescence intensity, which we take here to correspond to
the integrated area (IA) of the signal generated by an event\footnote{An event is the passage of one bead or cell through the flow cytometer. The MFI for  each event is calculated by measuring area under the curve for that event.}. The task of quantifying these effects can be understood as unfolding in two parts. First, we construct a model that quantifies how each source of
uncertainty propagates into the final measurement, which we take here to be \emph{raw integrated areas per event, not mean fluorescence intensities (or MFIs) of populations}. This ``forward process'' defines the final measurement uncertainty in terms of noise parameters associated with more granular physical effects.  Second, we treat the noise parameters as unknown and determine them as the output of a regression analysis that treats the  aforementioned model and real measurement data as inputs. This is the so-called inverse or backwards problem. In the two subsections that follow, we consider these steps in detail.

\subsection{The Forward Problem: Noise Modeling}

Our analysis assumes a population of objects containing fluorophores, such as  beads or cells.  For any given object, we generically denote the true underlying (and often unknown) number of flourophores by $f$.  The measurement of integrated area, to which we have access, is given by $x$.  Thus, the pair $(f,x)$ comprise the MESF-IA input-output relationship that we model.

Within this context, we model the measured signal $X$ as a random variable whose mean and variance are determined by autofluorescence, the random variable for number of fluorophores $F$, and other physical effects that we define below.  In particular, we posit that $x$, a particular realization of random variable $X$, can be generically expressed via the relationship \footnote{We will use $X,F$ to denote the random variables of measured signal and number of fluorophores and $x,f$ as the particular realizations of these random variables.}  
\begin{align}
x=\alpha f + \epsilon
\end{align}
where $\alpha$ is a multiplicative factor accounting for gain and sensitivity of the cytometer and $\epsilon$ is the associated noise term.  Different realizations $(f_1,\epsilon_1)$ and $(f_2,\epsilon_2)$ can clearly lead to the same value of $x$, and moreover for fixed values of $f$, different realizations of noise will lead to a distribution of $x$ values.  Thus, to determine the probability of a given value $x$, one needs to consider all the ways that $x$ could have been obtained from different underlying true signal values $f$, considering also the probability that the original number of fluorophores was $f$.  We therefore define the probability density $P(x)$ via the law of total probability \cite{ross2019first}.  This yields
\begin{equation}
P(x) = \int P(x | f) P(f) df
\label{eq: Model Total Prob}
\end{equation}
where the integral is over all possible $f$ values, and the notation $P(x|f)$ means the probability density of measurement $x$ conditioned on a fixed value of $f$.  Our task thus amounts to modeling the probability density $P(f)$ corresponding to distribution of measurands and the probability density $P(x|f)$ corresponding to measurement-induced variation associated with the instrument properties.

\subsubsection{Uncertainty in Measurands}

Consider different populations of measurands (e.g.\ beads), with each population being indexed by $j = \{0,1,,\cdots m\}$.  We assume that all populations have the same distribution of volumes; i.e.\ they come from the same manufacturing lot, for example.  Let the $j$th population have a nominal average \textit{volume} concentration $c_j$ of fluorophores.  That is, the $j$th population could correspond to the $j$th brightness level in a multi-peak bead set.  Let the number of events per population be $n_j$.  

In general, manufacturing and sample preparation processes often aim to produce measurands that are identical in size and fluorophore concentration.  However, members of a population always have inherent variation between them.\footnote{Note that if the beads are tagged on the surface instead of volumetrically, we need to consider surface area variation for tagging and volume variation for autofluorescence of the beads and the terms will change.}  We assume that the following unknown parameters characterize such effects: I. $\bar v$ and $\sigma_v$, which we take to be the mean and standard deviation of the volume; II. $\sigma_t$, which is associated with the standard deviation in fluorescence labeling (``t'' is for ``tagging''); and III. $k_a$, the autofluorescence per unit volume.   
Given these assumptions, we postulate that the true (unknown) $F_j$ are normal random variables of the form
\begin{subequations}
\begin{align}
F_j \sim N\left( \mu_{b_j}, \sigma_{b_j}^2 \right) &= (c_j + k_a) N_v(  \bar{v},  \sigma_v^2) + N_t(0, c_j\sigma_t^2) \label{eq:preq} \\
&=  N( (c_j + k_a) \bar{v}, (c_j + k_a)^2 \sigma_v^2 + c_j\sigma_t^2),
\label{eq: Model Q}
\end{align}
\end{subequations}
where $N_v$ and $N_t$ are independent normal random variables associated with volume and labeling variations, and $N$ is a normal random variable associated with a sum of the former two.  Note that Eq.\ \eqref{eq: Model Q} induces $P(f)$, which is simply the probability density function (PDF) of a Gaussian random variable.  

To justify the form of Eq.\ \eqref{eq: Model Q}, several comments are in order.  

First, the scaling of $\sigma_t^2$ by $c_j$ arises from the expectation that labeling processes follow Poisson or shot-noise statistics in the limit of many fluorophores.  In practice, this can correspond to as few as 50 to 100 average fluorophores per particle, since a Gaussian distribution still provides an adequate approximation \cite{ross2009introduction}. 

Second, observe that autofluorescence multiplies the volume variation term $\sigma_v^2$, but not $\sigma_t^2$.  This arises from the assumption that autofluorescence is an \textit{intrinsic} property of the material making up the particle, and thus depends solely on the volume.  In contrast, the nominal fluorophore concentration $c_j$ multiplies the additional term $\sigma_t^2c_j$, since we anticipate that the chemical reactions that achieve labeling have a degree of stochasticity independent of the volume.\footnote{Note that for surface-labeled beads or cells, the scaling of volume fluctuations by $c_j$ may change.  Moreover the term $c_j$ and $k_a$ may have different scalings from one another depend on whether the autofluorescence is due to bulk or surface effects.}  

Third, Eq.\ \eqref{eq: Model Q} amounts to a linear approximation insofar as we neglect effects due to simultaneous variations of labeling and volume.  That is, the \textit{nominal} concentration $c_j$ multiples $\sigma_v^2$, not its random counterpart depending on $N_t$.  Thus, we assume that products of the form $\sigma_v^2\sigma_t^2$ are sufficiently small as to be negligible.  See the Appendix for further details.

Finally, we note that in general, absolute number of fluorophores on a bead or cell is never known.  This is important because the variance in labeling should, in principle, be \textit{equal, not just proportional} to this number.  However, because we quantify concentration in arbitrary units of MESF, the quantity $\sigma_t^2$ effectively plays the role of converting between these units and the actual variance based on count.  For similar reasons, we may arbitrarily choose units associated with the nominal volume, which we set to $\bar v=1$ for convenience.  This amounts to a redimensionalization of all of the parameters appearing in Eq.\ \eqref{eq: Model Q}.  While it does not impact our final results, this rescaling is convenient from an optimization standpoint as it stabilizes subsequent computations.  We refer the reader to the Appendix for more details.  

\subsubsection{Instrument-Induced Uncertainty}

Given a model of $f$, we now turn to construction of $P(x|f)$.  An ideal instrument will measure the exact underlying signal.  That is, that is, the PDF $P(x)$ of integrated areas would simply equal $P(f)$, possibly up to a gain or scaling factor.  In this case, $P(x|f) = \delta(x-Gf)$, where $\delta$ is the Dirac delta function.  

This is, of course, not physically realistic, and the act of measurement itself changes the distribution of observed signal. In addition to scaling by the gain $G$, the measured signal $x$ is translated due to background $B$ before measurement, \footnote{Subsequent signals analysis often removes this \textit{on average} \cite{hoffman2007characterization}. However, this background still contributes to the measurement \textit{variance}. In such a scenario, instead of $B$ in both the terms, $B_1$ can be the background leftover in the mean and $B_2$ in the variance.}  Moreover, it is well known that photo-detector shot noise affects the signal, as well as recently studied gain-independent effects associated with Johnson-Nyquist noise \cite{kettlitz2014sensitivity,patrone2024uncertainty} .
This leads to a model of the form
\begin{equation}
X_j|(F_j=f_j) \sim N(G(f_j+B), G^2\sigma_s^2( f_j+B) + \omega),
\label{eq: Model X given Q}
\end{equation}
where $X_j$ is the random variable associated with a measurement of $F_j$, $\sigma_s^2$ quantifies the effect of photodetector shot-noise, and $\omega$ quantifies the gain-independent noise.  

It is important to note that Eq.\ \eqref{eq: Model X given Q} differs from more traditional ``Q and B'' models of cytometers, but only superficially so.  In particular, several authors postulate a sensitivity parameter $Q$ that quantifies the number of photoelectrons produced per incident photon on the detector \cite{hoffman2007characterization}.  In principle, this sensitivity can be estimated independently from background because shot noise scales linearly with $Q$ and quadratically in $G$; see Patrone et al \cite{patrone2024uncertainty} for an in-depth analysis.  However, in Sec.\ \ref{Section:Discussion} we show that Eq.\ \eqref{eq: Model X given Q} contains the same information as more traditional models as a result of rescaling units.  As a practical matter, we justify using Eq.\ \eqref{eq: Model X given Q} because it is more convenient to work with mathematically and as an object in optimization.    

\subsubsection{Approximation of Noise Model}

By combining Eqs.\ \eqref{eq: Model Total Prob},\eqref{eq: Model Q}, and \eqref{eq: Model X given Q}, we obtain an expression for the probability density function for the distribution of IA; see Eq. \eqref{eq:fullmodel} in Appendix.  In practice this expression is difficult to work with because it contains an integral that lacks a known antiderivative.  This is a consequence of the fact that the variance in Eq.\ \eqref{eq: Model X given Q} depends on the bead concentration $f_j$, which is itself random.  However, if we assume that the coefficients of variation of the $f_j$ are sufficiently small, we can approximate $f_j$ in Eq.\ \eqref{eq: Model X given Q} with its mean value.  While details are reserved for the Appendix, this yields the approximate model
%Such a model (), while more detailed, can be difficult to compute numerically. Under the assumptions that the beads distribution has a low coefficient of variation (CV) i.e. by assuming that the beads are all reasonably manufactured and tagged, we can approximate this distribution. We assume that the variance distribution can be approximated by the variance at bead mean $\mu_{b_j}$. By carrying this further (see Appendix), we obtain
\begin{subequations}
\begin{align}
X_j & \sim N(G\mu_{b_j},G^2 \sigma_{b_j}^2 + G^2 \sigma_s^2 (\mu_{b_j}+B)+\omega)\\
& \sim N(G(c_j \bar{v} + k_a\bar{v}),G^2 \left((c_j + k_a)^2 \sigma_V^2 + c_j\sigma_t^2\right) + G^2 \sigma_s^2 (B+c_j\bar{v}+k_a\bar{v})+\omega). \label{eq:Xj_approx}
\end{align}
\end{subequations}

In the next section, we further analyze this model and show how to determine all of the associated parameters given an appropriate calibration population.

%This approximation yields a much simpler distribution, easier to calculate while still including the crucial fact that the uncertainty of measurement inherently depends on the magnitude of the underlying true signal. We now use this approximate distribution from this point on.

\subsection{The Inverse Problem: Estimation of Parameters:}
\label{subsec:EstParams}

In a typical measurement setting, the quantities $\bm{\theta} =(c_0,c_1,...,c_m,k_a,\sigma_t^2,\sigma_v^2,\sigma_s^2,B)$ are unknown \textit{a priori} and must be determined through some calibration process.  The general procedure for estimating these parameters is to use a model in the spirit of Eq.\ \eqref{eq:Xj_approx} to construct an objective function $\mathcal L(\bm{\theta}|\mathcal{X})$, given some calibration data $\mathcal X$.  Minimizing this $\mathcal L$ then yields the desired parameter estimates.   

In this context, Eq.\ \eqref{eq:Xj_approx} is notable insofar as the concentration $c_j$ and autofluorescence $k_a$ always appear together as a linear combination \textit{except} in the term multiplying $\sigma_t^2$, which characterizes Poisson-noise in labeling.  It is straightforward to show that when the calibration data $\mathcal X$ is associated with beads for which $c_j$ is always unknown,\footnote{When considering calibration beads, it is important to note that the assigned MESF values are typically associated with the entire signal generated by an event, i.e.\ $G(c_j+k_a)$.  Thus, $k_a$ is unknown in this situation without extra information.} it is impossible to uniquely determine $k_a$, $c_j$, $\sigma_v^2$, and $\sigma_t^2$.  While mathematical details are provided in the Appendix, this observation can be intuitively justified on physical grounds.  For example, note that on average, dim but highly autofluoresceing beads yield the same measurements as bright, non autofluoresceing beads.  In fact, the resulting measurement \textit{distributions} can be made equal, provide $\sigma_t^2$ is rescaled appropriately.  

The solution to this problem is to assume that the calibration data includes measurements from blank beads having concentration $c_0=0$ and the same autofluorescence $k_a$ as all other populations.  This data allows one to effectively separates $\sigma_v^2$ and $\sigma_t^2$ because autofluorescence is a property of the particle substrate, and thus only scales with its mass, i.e.\ volume; see the Appendix.  As an aside, we caution the reader that obtaining blank beads with the same autofluorescence as the other populations is not as trivial as it sounds.  Even for multipeak calibration sets, we have found that manufacturers do not always pull blank beads from the same lot as the labeled beads.  See the SI of Ref.\ \cite{patrone2024uncertainty}, which illustrates this problem, as well as a means for detecting blank beads with distinct $k_a$.  

Given these observations, let $\bm{\theta} = [B\  \sigma_s^2\  \textbf{c}\  k_a\  \sigma_V^2\  \sigma_t^2]$, where $\textbf{c} = [c_0,c_1,\cdots,c_m]$ are the nominal average tagged fluorophores per unit volumes for different beadtypes.  As a practical matter, we can set $\bar v=1$ in the computations and rescale all quantities after-the-fact if needed.  Moreover, let $\mathcal X = \{\mathcal X_j\}$ denote the set of measurements, where $\mathcal{X}_j=\{x_{i,j}:i=1,\cdots,n_j\}$ for beads of type $j\in \mathcal{J} = \{0,1,2,\cdots,m\}$, and denote the set of measured gains by $\mathcal{G}$.  We define the objective function to be the negative log likelihood function  \cite{ross2009introduction}
\begin{equation}
\mathcal{L}(\bm{\theta}|\mathcal{X})
 = \sum_{G\in \mathcal{G}, j\in \mathcal{J}} \left( \frac{n_j}{2} \ln(2\pi)+ \frac{n_j}{2} \ln(G^2 \sigma_{b_j}^2 + G^2 \sigma_s^2 (\mu_{b_j}+B)+\omega) + \frac{1}{2} \sum\limits_{i=1}^n \frac{(x_{i,j}-G(B+\mu_{b_j}))^2}{G^2 \sigma_{b_j}^2 + G^2 \sigma_s^2 (\mu_{b_j}+B)+\omega}\right).
\end{equation}
To estimate the $\theta$, we minimize $\mathcal L(\bm{\theta}|\mathcal X)$ as a function of these parameters.  See the Appendix for more details.%We minimize this simultaneously for all beads to obtain an estimate of parameters. Notice that there is manifold on which this objective function has constant value, see Appendix where this manifold is explicitly described. However, if untagged (blank) beads are part of the bead sets included and if the measurements are repeated on multiple gains, we show that this manifold is resolved and various parameters involved can be determined. Any orthogonal measurements of the physical properties of the beads can help strengthen the resolution further.

\begin{figure}[!htbp]
\centering
\includegraphics[width=0.8\textwidth]{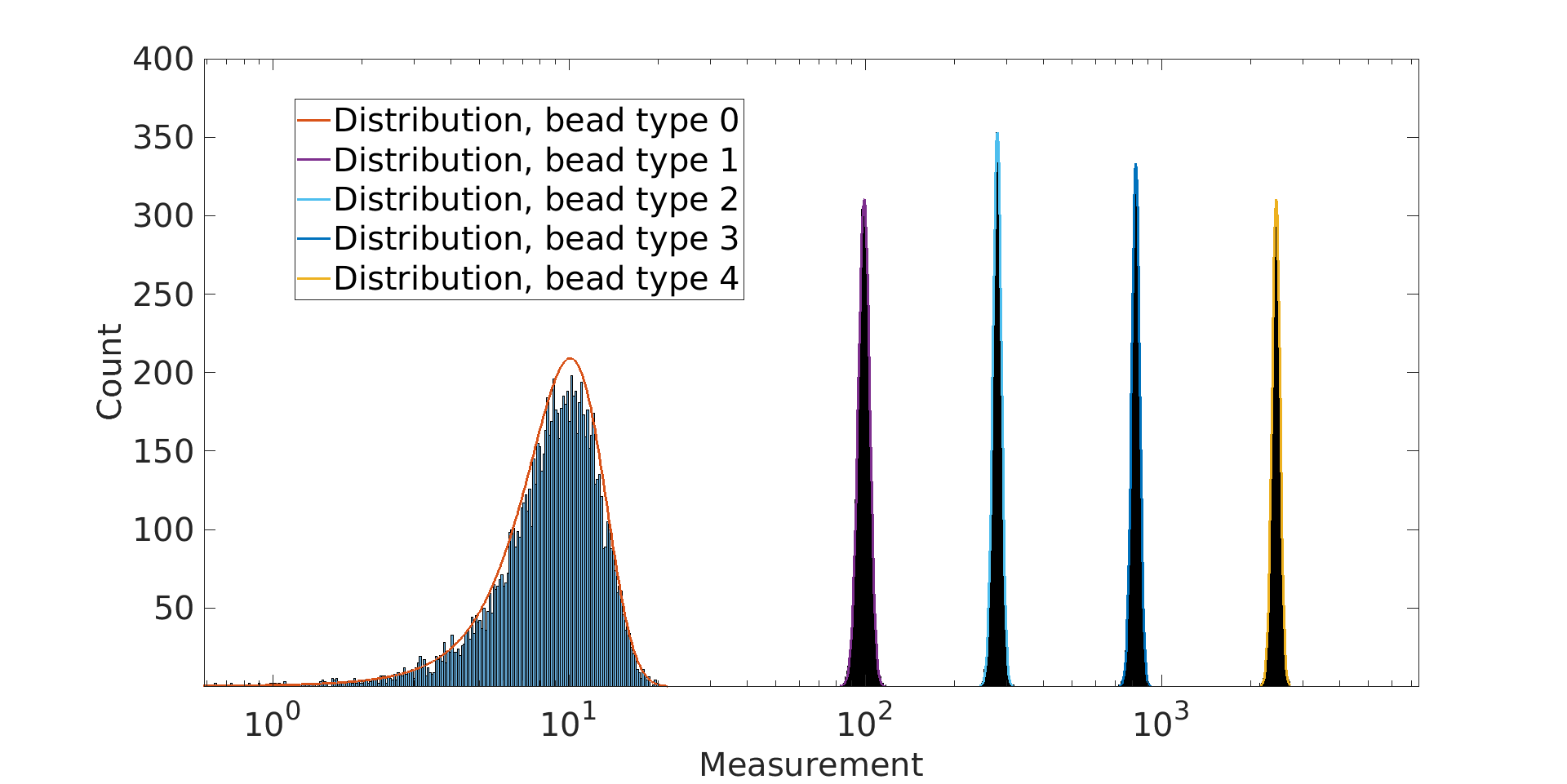}
\caption{Logscale histogram and the corresponding distribution for synthetic data, all bead types. Bead type 0 corresponds to unlabeled (blank) beads. relative gain = $9$ units. Although histograms are shown for illustrative purposes, the model is fit in a histogram-free fashion, using only per event data. x-axis is measurement with the units of MFI.}
\label{fig:gain9_synth_log}
\end{figure}

\begin{table}[!htbp]
\[\begin{array}{|c|c|c|}
\hline
\text{Parameter} & \text{True value} & \text{Estimated value}\\
\hline
B & 1 & 1.0087\\
\sigma_s^2 & 0.001 & 0.0009\\
k_a & 0.01 & 0.0011\\
\sigma_V^2 & 0.001 & 0.001 \\
\sigma_t^2 & 0.001 & 0.0008\\
c_1 & 10 & 9.9974\\
c_2 & 30 & 30.0001 \\
c_3 & 90 & 89.9997 \\
c_4 & 270 & 270.0097\\
\omega & 10 & 10.0714\\
\hline
\end{array}\]
\caption{True parameter values for synthetic data alongside estimated parameters, variation in the 3rd decimal. Estimates come from method detailed in subsection \ref{subsec:EstParams} applied to the synthetic data.}
\label{table:synth}
\end{table}

\begin{table}[!htbp]
  \centering
\[\begin{array}{|c|c|c|}
\hline
\text{Bead type} & \text{Percent error in mean} & \text{Percent error in variance}\\
\hline
0 & 0.184 & 0.276 \\
1 & 0.004 & 0.567\\
2 & 0.020 & 2.680\\
3 & 0.001 & 0.595\\
4 & 0.044 & 1.608\\
\hline
\end{array}\]
\caption{Percent errors in mean and variance of measurements for synthetic data at relative gain = 9 units. The mean and variance for the model were calculated using equation \eqref{eq:Xj_approx}.}
\label{table:synth_means_manif}
\end{table}

\subsection{Data Acquisition}
\label{Section:data_acquisition}

Samples containing MESF-specified polystyrene microspheres (9 intensity peak ultra rainbow beads with nominal mean diameter $\approx$ 10.2~µm, URCP-100-2H, AP03, Spherotech\footnote{The full description of the procedures used in this paper requires the identification of certain commercial products and their suppliers. The inclusion of such information should in no way be construed as indicating that such products or suppliers are endorsed by the National Institute of Standards and Technology (NIST) or are recommended by NIST or that they are necessarily the best materials, instruments, or suppliers for the purposes described.}, Lake Forest, IL) at a concentration of $2\times 10^6$ beads per mL were prepared in filtered water mixed with saturated salt at a volume fraction of approximately 25~\% (for neutral buoyancy of the beads) and with a surfactant, Triton-X 100, at a volume fraction of 0.1~\% (to prevent bead aggregation). The fluorescence intensity in the FITC channel (emission bandwidth: 525~nm $\pm$ 20~nm) from the sample was measured on a commercial cytometer (50~mW laser excitation at 488~nm with avalanche photodiodes for fluorescence and silicon photodiode for forward scatter (FSC) detection). All events were collected using the slowest sample flow rate, which was 10 $\mu$L per min ($\approx 1.6667 \times 10^{10}~m^3$/s) from peristaltic pumps.
 The cytometer detector's linear gain was swept in factors of three to sweep a broad range (i.e. 30, 90, 270, 810). Bead populations on the commercial cytometer were gated using all available fluorescence and FSC channel metrics after dimensionality reduction to two dimensions via principal component analysis (PCA). Per-event area under the curve of the signal in the FITC- channel (FITC-A or IA) for each bead was recorded and used for further calculations.

\FloatBarrier
\section{Results}
\label{Section:Results}

To demonstrate our method, we first test it on synthetic data generated with known parameters. These datasets are generated with relative gains $\mathcal{G} = \{1,3,9\}$ for unlabeled beads (bead type 0) as well as four sets of beads (bead types 1 to 4) with mean tagged relative intensities $10,30,90,270$ respectively. Other details about the true parameters and the estimated quantities are in Table \ref{table:synth} \hspace{-0.3cm}.
The data histograms and corresponding distributions are displayed in Figure \ref{fig:gain9_synth_log} \hspace{-0.3cm}. The percent error of means and variances for these estimates compared to the underlying data is in Table \ref{table:synth_means_manif} \hspace{-0.3cm}. Notice that the estimate has captured the underlying statistics of the data very well, differing only at the third decimal place.

\begin{figure}[!htbp]
\centering
\includegraphics[width=0.8\textwidth]{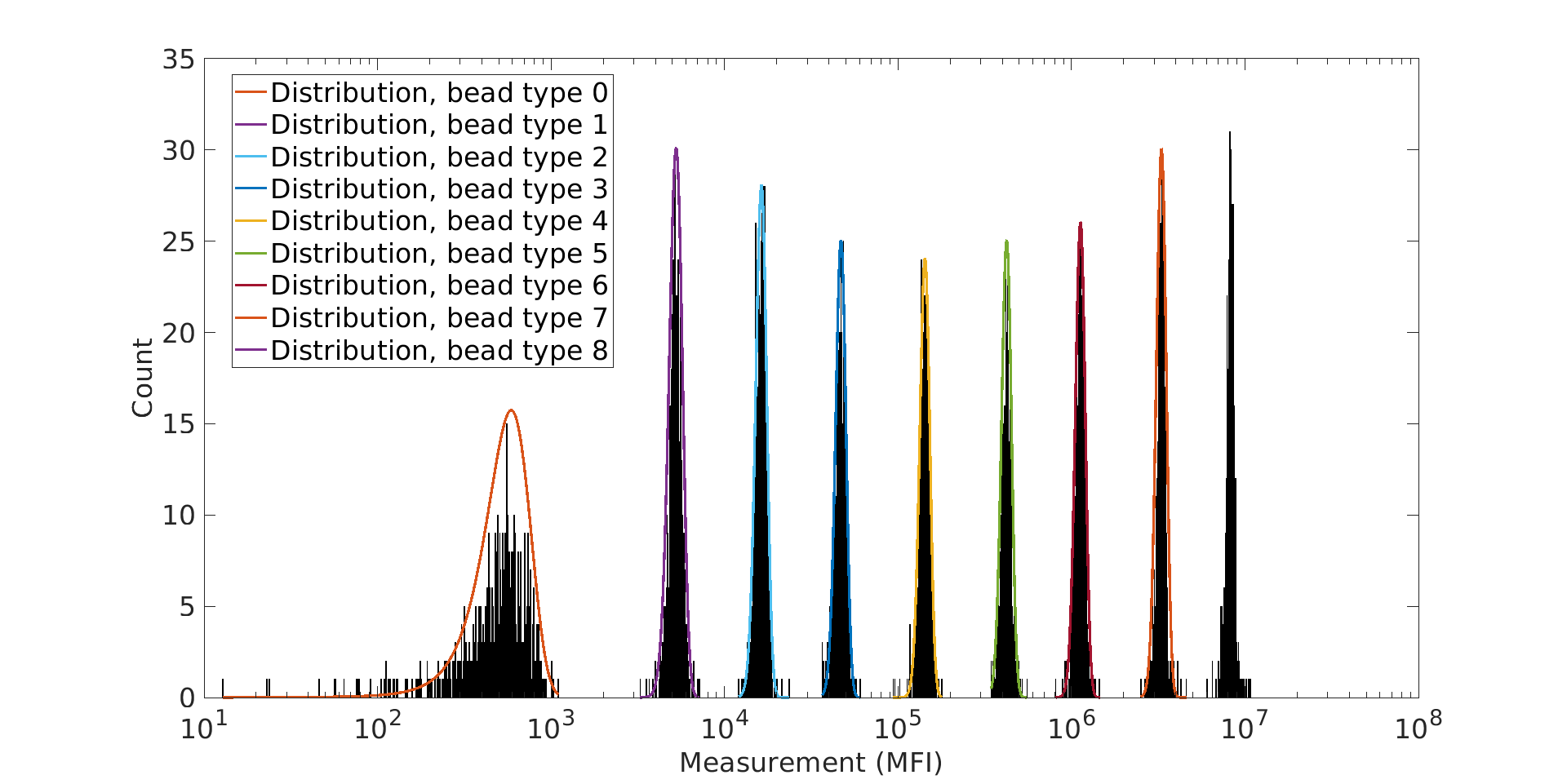}
\caption{ Logscale histogram and the corresponding distribution for real data, all beadtypes. Bead type 0 corresponds to unlabeled (blank) beads, gain = $90$ units. Although histograms are shown for illustrative purposes, the model is fit in a histogram-free fashion, using only per event data.}
\label{fig:gain90_data_log}
\end{figure}

\begin{table}
\[\begin{array}{|c|c|}
\hline
\text{Parameter} & \text{Estimated values}\\
\hline
B & 5.8120 \\
\sigma_s^2 & 0.1100 \\
k_a & 0.0350 \\
\sigma_V^2 & 0.0044\\
\sigma_t^2 & 0.1015\\
c_1 & 52.3001\\
c_2 & 174.0296\\
c_3 & 514.9349\\
c_4 & 1.5782 \times 10^3\\
c_5 & 0.4702 \times 10^4\\
c_6 & 1.2415 \times 10^4\\
c_7 & 3.3158 \times 10^4\\
c_8 & 8.5245 \times 10^4\\
\omega & 2.5301\times 10^4\\
\hline
\end{array}\]
\caption{Estimated parameter values for real data, further explanation of the parameters in section \ref{Section:Theory} \hspace{-0.3cm}. The model fit here is done in a histogram-free fashion, using only per-event data. The derived intensities $c_i$ scale with a factor of three as expected without any additional constraints. Estimates come from method detailed in subsection \ref{subsec:EstParams} applied to the real data.}
\label{table:real}
\end{table}

\begin{figure}
\centering
\begin{subfigure}[t]{.48\textwidth}
  \centering
  \includegraphics[width=\textwidth]{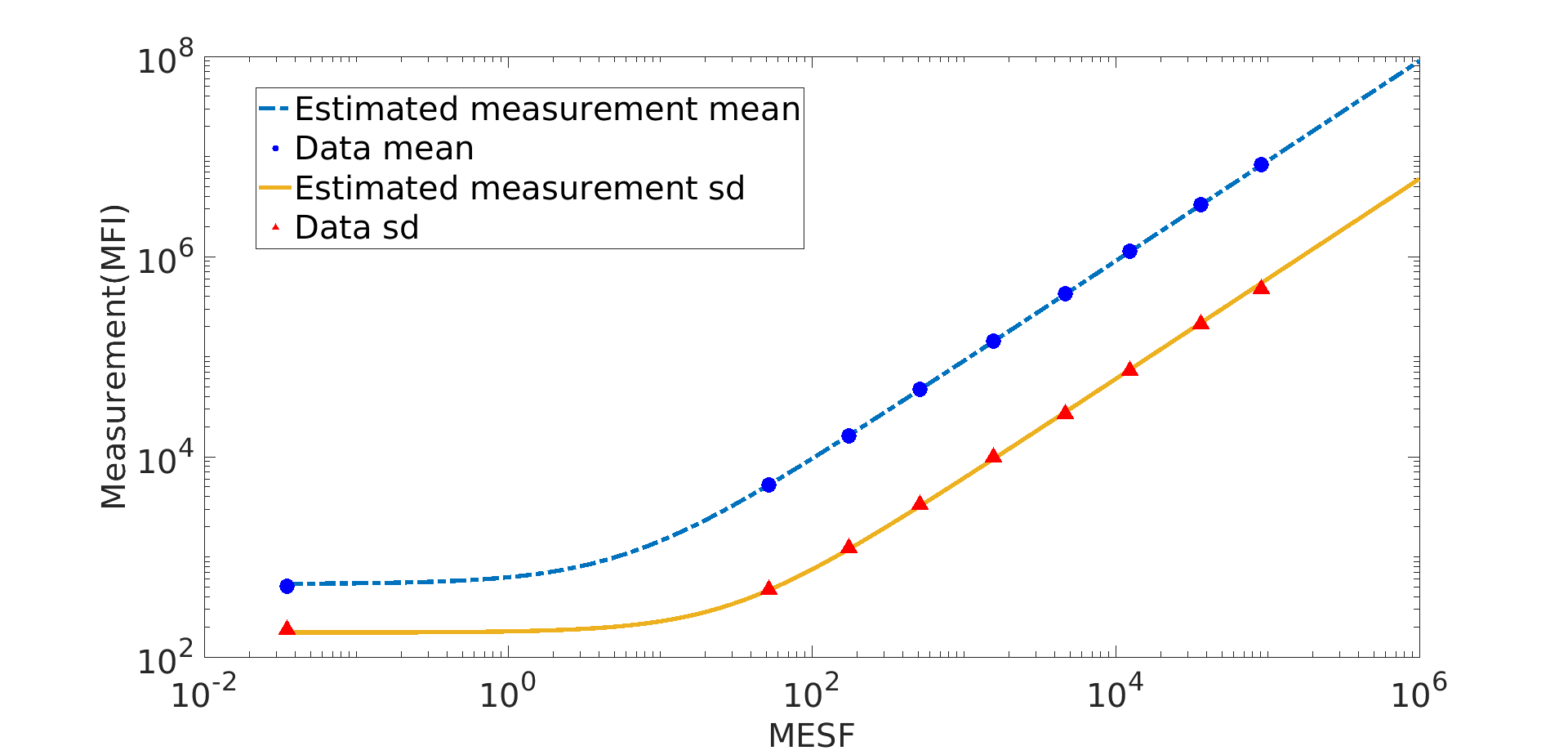}
  \caption{Estimated curve for empirical measurement mean and standard deviation (sd) for true intensities and corresponding bead-data measurement. Gain = $90$ units. The $95\%$ Confidence Intervals on empirical means and sd are too narrow to be displayed effectively.} 
  \label{fig:gain90_mean_sd}
\end{subfigure}\hfill%
\begin{subfigure}[t]{.48\textwidth}
  \centering
  \includegraphics[width=\textwidth]{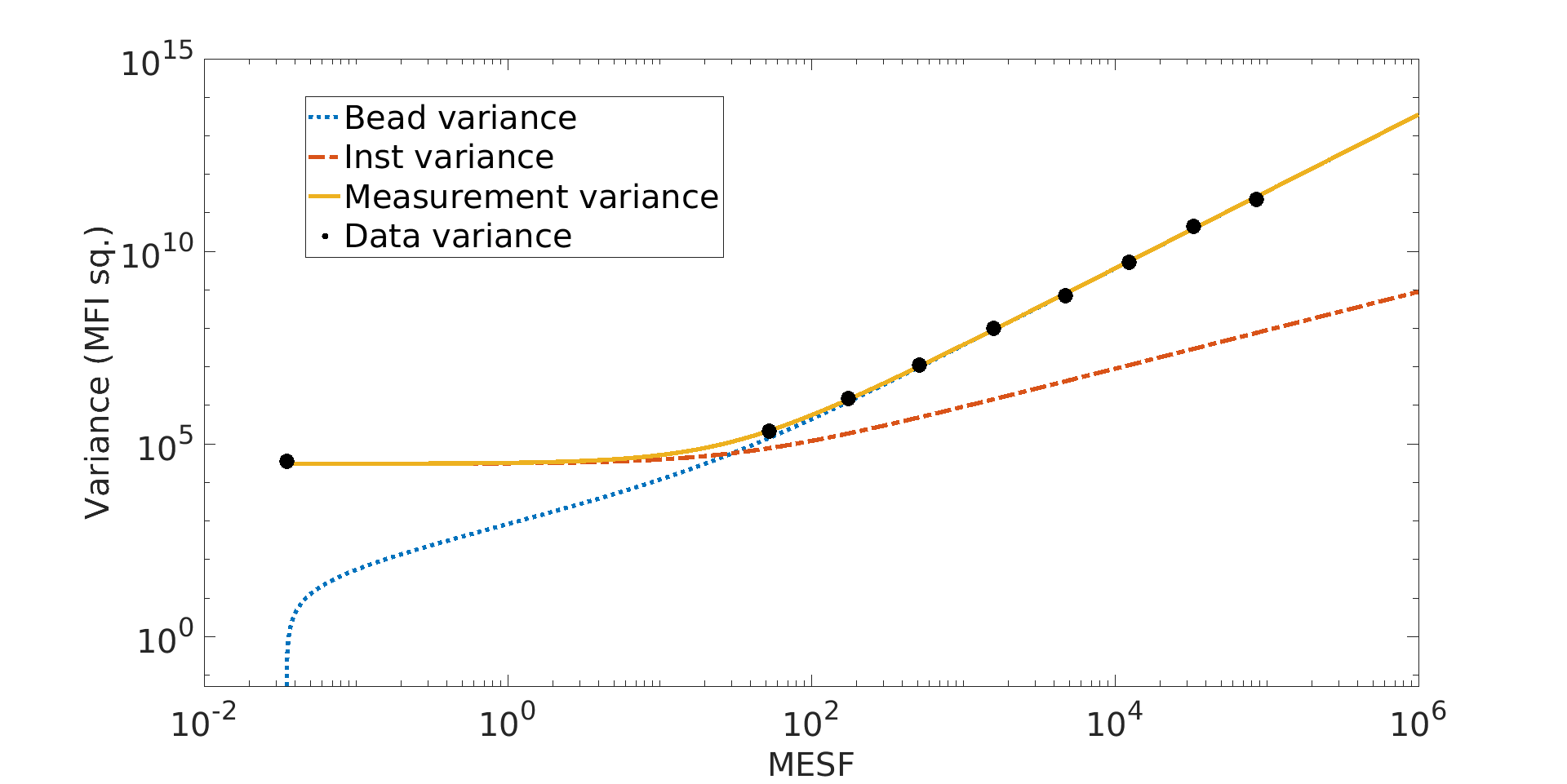}
  \caption{Empirical and model variance and corresponding uncertainty contributions from beads and instruments. Gain = $90$ units.}
  \label{fig:sd_gain90_data}
\end{subfigure}
\caption{Uncertainty decomposition using estimated parameters and real experimental data. The model is fit in a histogram-free fashion, using only per event data.}
\label{fig:uncer_decomp}
\end{figure}

\begin{figure}
\centering
\begin{subfigure}[t]{.48\textwidth}
  \centering
  \includegraphics[width=\textwidth]{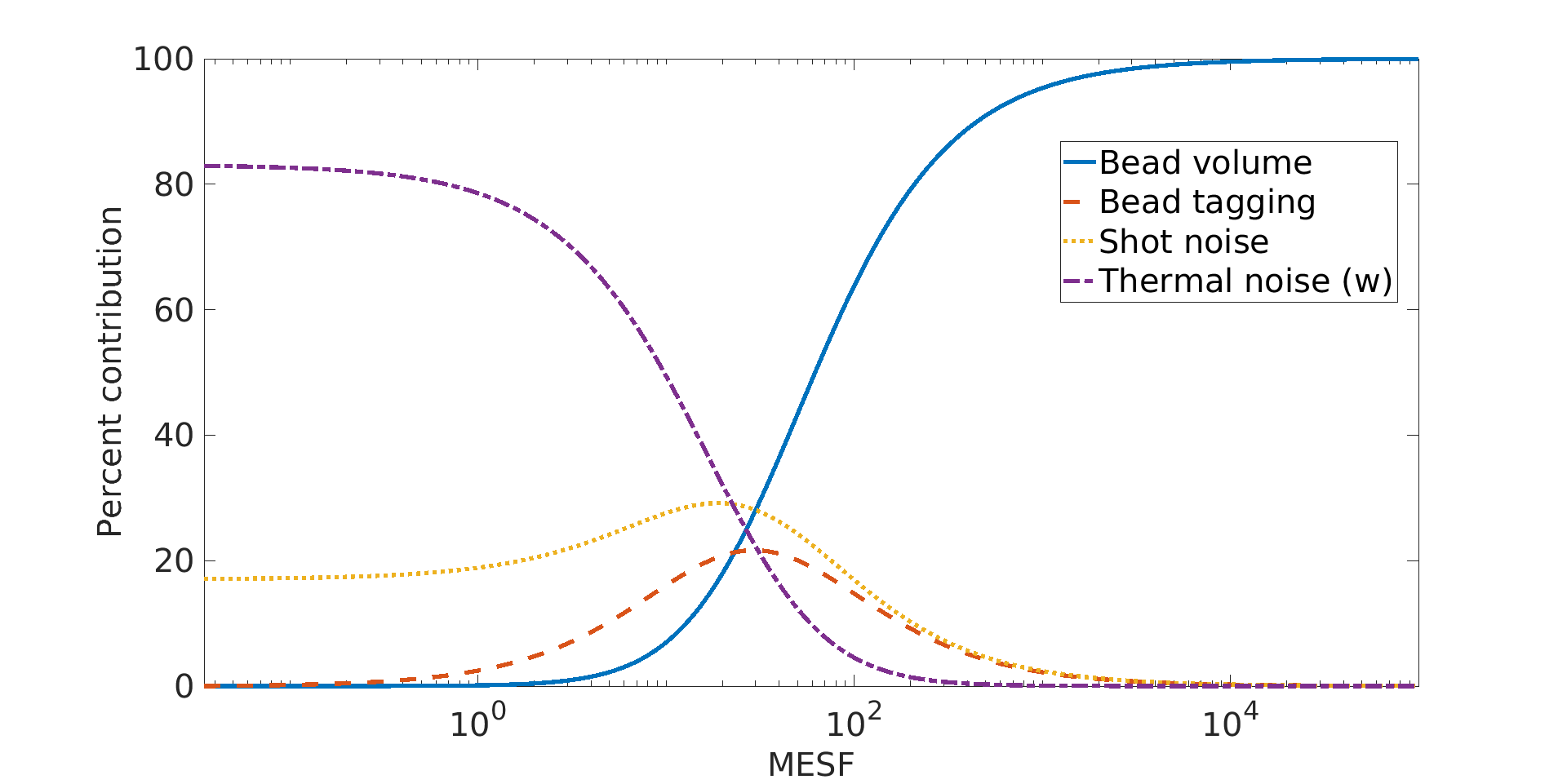}
  \caption{Percent contributions of different variances to measurement variance for data. Gain = $90$ units.}
  \label{fig:data_variance_percentcontr_3}
\end{subfigure}\hfill%
\begin{subfigure}[t]{.48\textwidth}
  \centering
  \includegraphics[width=\textwidth]{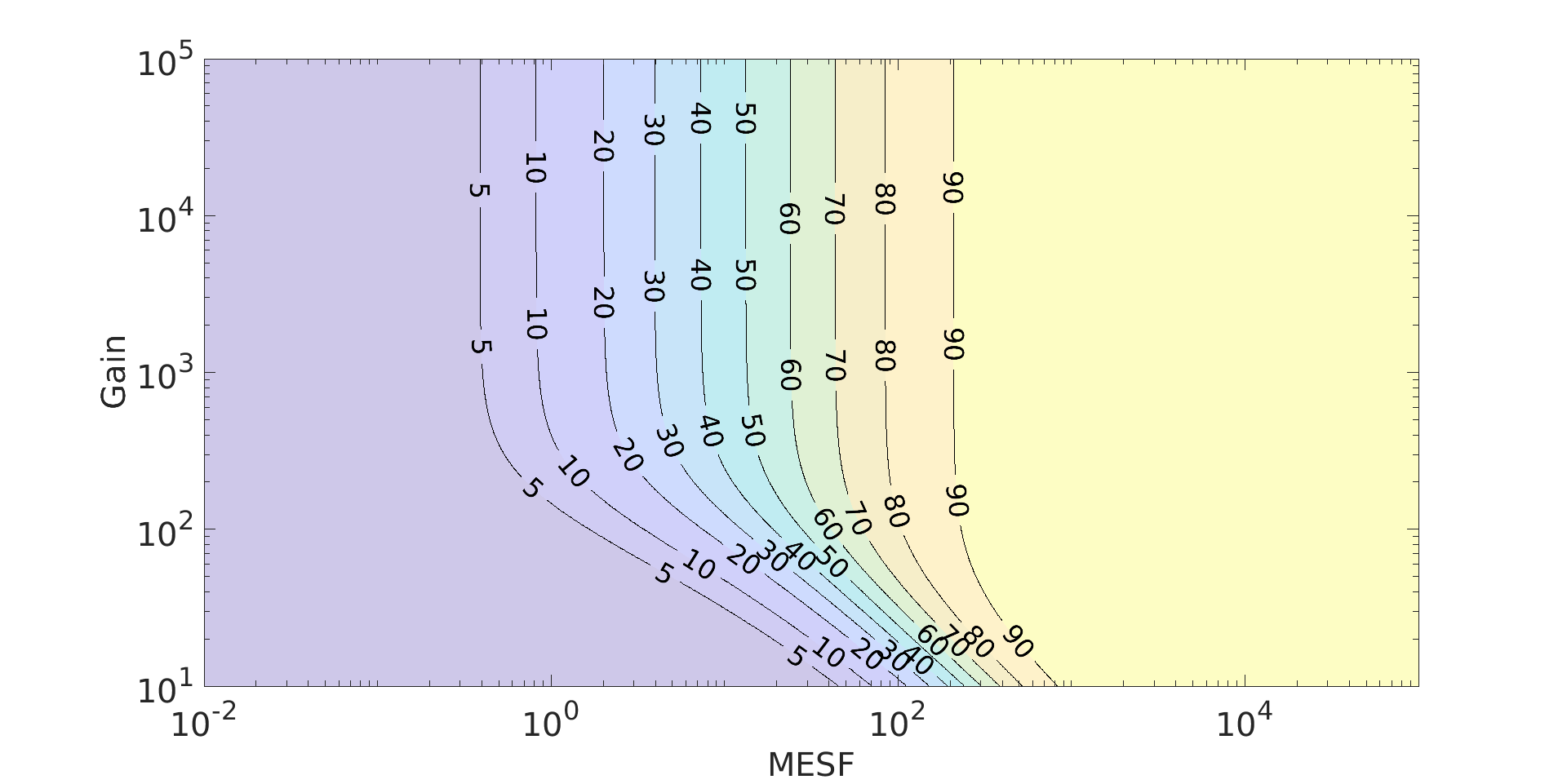}
  \caption{Plot for percent contribution by bead variance for measurement variance over a range of MESF and gains, calculated using the estimated parameters.}
  \label{fig:mesf_gain_thetasol_beadpercent}
\end{subfigure}
\caption{Percent contribution to uncertainty using estimated parameters}
\label{fig:percent_decomp}
\end{figure}

In order to validate the analysis on real data, we used commercially available fluorescence calibration beads with 9 different intensities but the same size (nominal average diameter of $\approx 10.2 ~\mu$m).  See Section \ref{Section:data_acquisition} for more details of the underlying measurements.  As before, maximum likelihood estimation was applied to this data. Bead type $0$ are unlabeled, and as such, the resulting signals were due to a combination of autofluorescence and instrument uncertainties. This is implemented by fixing $c_0=0$. The data as well as the distributions associated with the model are shown in Figure \ref{fig:gain90_data_log} \hspace{-0.3cm}. The corresponding parameter estimates are displayed in Table \ref{table:real} \hspace{-0.3cm}.  Note that these parameters were obtained with the assumption that $\bar{v} = 1$; rescaling can be performed after-the-fact to obtain parameter estimates with appropriate units. Background $B$ is a more significant source of uncertainty in this case than the autofluorescence $k_a$.  Of note, the scaling of MESF values between beads automatically falls out of the analysis, as seen in the $c_i$ values.

Figure \ref{fig:gain90_mean_sd} illustrates how the empirical mean and standard deviation (SD) of each population compare to the predicted mean and SD arising from the model.  %We can assign these estimated mean and standard deviations to even the MESF values for which there was no bead available. 
Figure \ref{fig:sd_gain90_data} shows individual contributions to the variance arising from the beads and instrument. In low intensity regions, the contribution from beads is much lower than the corresponding contribution from the instrument. In Figure \ref{fig:data_variance_percentcontr_3}, it is shown that  large percent of this contribution comes from gain-independent instrument uncertainty, and a large percent of contribution at higher MESF comes from bead volume variance. Percent contributions are plotted for different gains and MESF in Figure \ref{fig:mesf_gain_thetasol_beadpercent} \hspace{-0.1cm}; we notice that the MESF at which different effects dominate change with gain, albeit the general trend is the same across gains.  In fact, the qualitative behavior should be largely independent of the beads or instrument used.  Refer to subsection \ref{subsec:trends_uncert} for further details.

\textit{\textbf{Noise-to-signal ratio (NSR)}} is defined as the standard deviation of a population divided by its mean. It is a commonly used metric to identify population variability in relation with the underlying signal. In context of our model,

\begin{equation}
NSR_{\text{meas}} = \frac{\sqrt{G^2 \left((c + k_a)^2 \sigma_V^2 + c\sigma_t^2\right) + G^2 \sigma_s^2 (B+c+k_a)+\omega}}{G(B+c  + k_a)}.
\end{equation}

\begin{figure}
\centering
\begin{subfigure}[t]{.48\textwidth}
  \centering
\includegraphics[width=\textwidth]{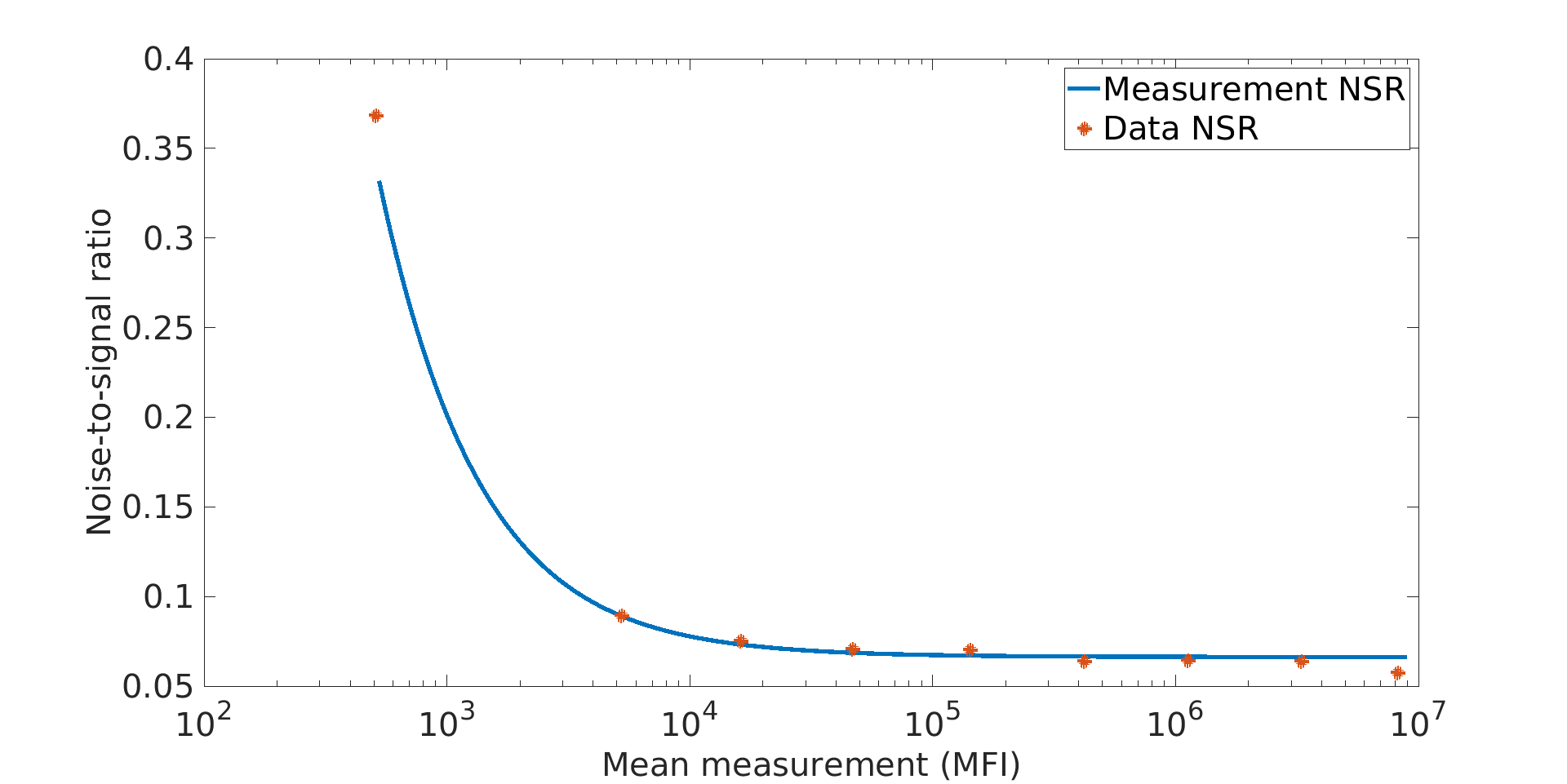}
\caption{Plot for noise-to-signal ratio dependent on mean measurement values (MFI), calculated using the estimated parameters and with data, gain = 90 units.}
\label{fig:meas_cv}
\end{subfigure}\hfill%
\begin{subfigure}[t]{.48\textwidth}
  \centering
\includegraphics[width=\textwidth]{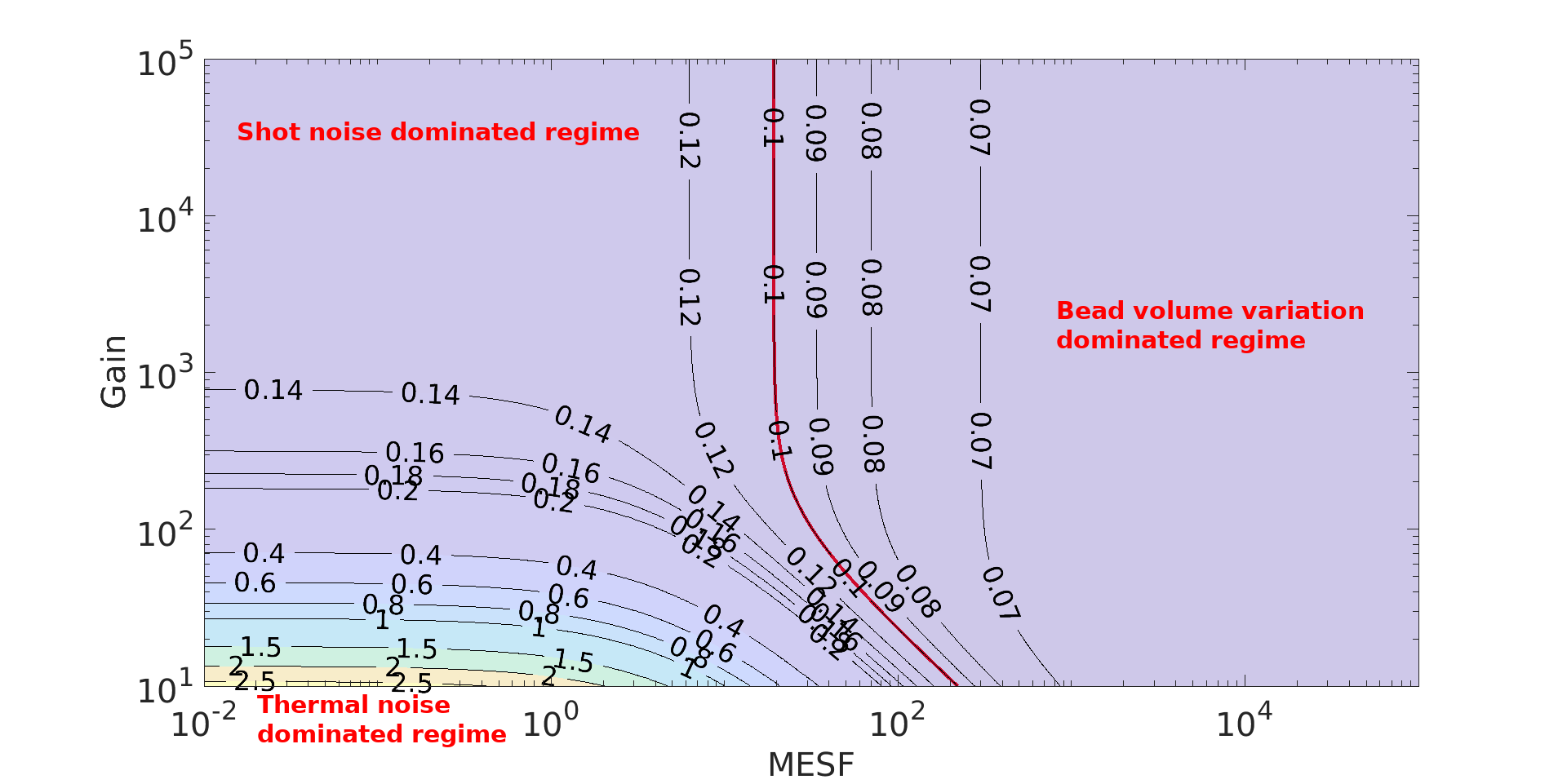}
\caption{Plot for noise-to-signal ratio for a range of MESF and gains, calculated using the estimated parameters.}
\label{fig:mesf_gain_thetasol_nsr}
\end{subfigure}
\caption{Noise-to-signal ratios}
\label{fig:nsr}
\end{figure}

Figure \ref{fig:meas_cv} shows the agreement between the data and the estimates from the model for noise-to-signal ratio. Notice that this is in terms of MFI and only displayed here for a single gain. In order to ensure meaningful decision-making, we also report how the noise-to-signal ratio changes with gain and MESF, Figure \ref{fig:mesf_gain_thetasol_nsr}. If the beads used had similar population CV as the cell population to be measured, we can use this plot to identify reasonable gain settings to run our sample. Refer to subsection \ref{subsec:uq_calibration} for more details.

\section{Discussion} 
\label{Section:Discussion}

\subsection{Comparison to Existing Methods}

To better highlight the novel aspects of our analysis, it is useful to frame them the context of previous works.  In particular, many authors have considered so-called $Q$ and $B$ or ``sensitivity'' and background analyses \cite{hoffman2007characterization,chase1998resolution}.  Historically this perspective has been used  to characterize what are essentially optical properties of the system.  However, to the best of our knowledge, such methods are largely detached from deeper metrology concepts such as noise-to-signal ratio.  An exception to this is the recent work \cite{patrone2024uncertainty,catterton2024uncertainty}, but even the results therein do not permit conventional cytometers to distinguish instrument and measurand-based sources of uncertainty.

From a mathematical standpoint, this discrepancy appears to arise largely from the way in which past works have formulated models and accounted for uncertainty, not a fundamental difference in the underlying mathematics.  Recourse to Eq.\ \eqref{eq: Model Total Prob} explains this fact.  Note, in particular, that even when Eq.\ \eqref{eq: Model Total Prob} cannot be computed exactly, we can still often compute moments of the form
\begin{subequations}
\begin{align}
\langle x \rangle &= \int x P(x|f)P(f)\, df dx, \\ 
\left \langle \left (x - \langle x \rangle \right )^2 \right \rangle &= \int  \left (x - \langle x \rangle \right )^2 P(x|f)P(f)\, df dx.
\end{align}
\end{subequations}
In fact, it is straightforward to show that for our model, 
\begin{subequations}
\begin{align}
\langle x_j \rangle &= G(c_j + k_a + B), \label{eq:exactcmean} \\ 
\left \langle \left (x_j - \langle x_j \rangle \right )^2 \right \rangle &= G^2 \left((c_j + k_a)^2 \sigma_V^2 + c_j\sigma_t^2\right) + G^2 \sigma_s^2 (B+c_j\bar{v}+k_a\bar{v})+\omega \\
&= G^2\sigma_v^2(c_j + k_a)^2 + G^2(\sigma_t^2 + \sigma_s^2 \bar v) (c_j+k_a) + G^2(\sigma_s^2B - k_a \sigma_t^2) + \omega, \label{eq:exactcvar}
\end{align}
\end{subequations}
even when using the exact expressions for $P(x|q)$ and $P(q)$.  Equation \eqref{eq:exactcvar} reveals that the variance is a quadratic function $\varsigma_j^2 = \alpha + \beta (c_j+k_a) + \gamma (c_j+k_a)^2$.  While this is consistent with models considered in past works \cite{hoffman2007characterization}, those works do not \textit{derive} the coefficients $\alpha$, $\beta$, and $\gamma$ from the physics of a more granular model, instead assigning them a single phenomenological interpretation.  Thus, while our analysis justifies the prior use of quadratic models, it also highlights that previous works conflate the interpretations of $\sigma_t^2$ and $\sigma_s^2$, and potentially other terms, depending on whether or not autofluorescence is grouped with $c_j$ in the calibration measurements.  

To make this more precise, we consider the analysis of Ref.\cite{chase1998resolution} and group $k_a$ with $c_j$, meaning that both are considered to contribute to the measured MESF.  The authors of that work assume that the mean fluorescence intensity (MFI) and variance in fluorescence measurements associated with a population can be expressed in the form
\begin{align}
\langle x_j \rangle &= \hat G Q (c_j + k_a) \label{eq:oldqandbmean} \\
\left \langle \left (x_j - \langle x_j \rangle \right )^2 \right \rangle&= \hat G^2 Q \hat B + \hat G^2 Q (c_j + k_a) + \hat G^2 Q^2 \varsigma^2 (c_j + k_a)^2, \label{eq:oldqandbvar}
\end{align}
where $\hat G$ is their gain factor, $Q$ is the sensitivity of the photodetector, $\hat B$ is a background source, and $\varsigma^2$ is an unspecified variance \cite{patrone2024uncertainty}.\footnote{In Ref.\cite{chase1998resolution}, the term proportional to $\varsigma^2$ was only implicitly assumed in the model.  For more details and justification on this observation, see the discussion of Ref.\ \cite{patrone2024uncertainty} }  Comparing Eqs.\ \eqref{eq:oldqandbmean} and \eqref{eq:oldqandbvar} to Eqs.\ \eqref{eq:exactcmean} and \eqref{eq:exactcvar} and momentarily ignoring $\omega$ (whose significance was only recently uncovered), the substitution $G = \hat G Q$ implies that 
\begin{subequations}
\begin{align}
\varsigma^2 &= \sigma_v^2,\\
\frac{1}{Q}&= \sigma_t^2 + \sigma_s^2 \bar v, \\
\frac{\hat B}{Q} &= \sigma_s^2B - k_a \sigma_t^2.
\end{align}
\end{subequations}
If the labeling variation $\sigma_t^2$ is zero, we immediately recover that $\sigma_s^2 = Q^{-1}$ and $\hat B = B$.  The first of these in particular is physically insightful: the lower the sensitivity of the photodetector, the larger the shot noise, since fewer photons are converted to photoelectrons.  \textit{However}, if the shot noise due to labeling is not small (as could happen for dim particles), the background $\hat B$ and sensitivity $Q$ becomes conflated with autofluorescence and labeling variation.

The utility of these observations ultimately depend on the information that one wishes to extract from cytometry measurements, as well as the systems to which this analysis can be applied.  In the following sections, we consider these issues in more detail.

\subsection{Extensions to Cell-Based Measurements}

A surprising aspect of the analysis presented herein is its ability to quantify the relative concentration of fluorophores in different populations without knowledge of the underlying MESF values associated with the sample.  In particular, the analysis treats the concentrations $c_j$ as unknowns to determined via optimization, with the exception of $c_0=0$, which is associated with unlabeled beads.  In so doing, we also obtain estimates of autofluorescence $k_a$.

From a conceptual standpoint, this observation means that our analysis is not restricted to beads \textit{per-se}, and an important open question is the extent to which Eqs.\ \eqref{eq: Model Q} and \eqref{eq: Model X given Q} are valid for such systems.  From a practical standpoint, running the necessary experiments on cells would entail measuring unlabeled samples, as well as those with differing concentrations of labeling.  Provided this is possible, the analysis should be able to estimate not only the autofluorescence of cells, but also distinguish volume and labeling variation.  

We anticipate that this latter feature is especially important because it detangles distinct biological phenomena that contribute to the measurement process in a similar way.  In particular, consider that cells can be expected to have variations in both their size and number of biomarkers of interest.  In the latter case, we speculate that the labeling variation $\sigma_t^2$ can be tied to the variation in concentration of these biomarkers.  This could provide two ``orthogonal'' views of population distributions by determining whether the spread in measurements of a cell population is due primarily to heterogeneity in size, biomarker abundance, or both.  However, we note that such an analysis would likely need to take into account for sample preparation, reagent handling, and the thermodynamics of antibody binding (if such macromolecules are used for labeling).  Moreover, the assumption of small CV used herein may not be valid, which would require use of the full model, not approximations thereof.  While such issues are left for future work, we anticipate that extensions of the analysis herein could address such problems.

\subsection{Uncertainty Quantification on a Per-Event Basis}
\label{subsec:uq_calibration}
\label{subsec:trends_uncert}

An advantage of the modeling approach considered herein is the ability to both quantify uncertainty on a \textit{per-event} basis and identify the dominant contributions to that uncertainty.  We anticipate that these benefits can address at least two important problems when performing cytometry measurements.  

First, the results of our analysis allow one to determine reasonable settings for performing measurements.  Consider, for example, Figure \ref{fig:mesf_gain_thetasol_nsr}.  Both subfigures show variations on the noise-to-signal (NSR).  Given two populations that one wishes to distinguish, this figure can be used to determine, for example, reasonable gain settings that minimize the excess noise due to the measurement process.  

Perhaps more importantly, our analysis may also allow one to determine \textit{when} a given effect is dominating the measurement uncertainty and thereby suggest mitigation strategies; see, for example, Figure \ref{fig:percent_decomp}.  Considering cell-based measurements, for example,  a large value of $\sigma_t^2$ even when the expected biomarker concentration is large suggests a need to revisit the labeling process and reagents used in sample preparation.  A variation on this idea was proposed in a previous study \cite{wang2021establishing}, but that work required several repetitions of the sample preparation process in an effort to characterize its impact.  Thus, having the ability to quantify the associated uncertainties without creating new samples would provide a more efficient route for addressing this problem.  

Finally, we note that our analysis provides an orthogonal, partly indirect characterization of standards such as variation in beads. For instance, for the real data used, the bead specification sheet provides a mean diameter of $\approx 10.2~\mu$m with a range of $8.0~ \mu$m$-12.9~\mu$m. We find however that this interval can be substantially improved by using our method. We obtain a standard deviation of $0.2248~\mu$m. The 3 standard deviation interval around the mean thus amounts to a $[9.52, 10.87] ~\mu$m interval for the beads. Further verification by physical measurements can help improve the estimates.

\subsection{Limitations}

While we have demonstrated that our approach subsumes many past works, our model still makes certain assumptions about bead distribution that may not be true in general. For instance, if the beads are tagged on the surface instead of volumetrically, the model will have to be reformulated to account for surface variance for tagging and  volume variance for autofluorescence. There is also an underlying assumption of instrument-dependent constant background and gain-independent part of the variance which might not necessarily hold true unless the calibration is done immediately before the sample measurements. Many instrument effects such as laser profile, optical path of the beads, flow rate, etc were not separately considered, however, research has shown \cite{gaigalas2021sources} that some of these factors do not have a large effect on measurement uncertainty provided  best-practice guidelines are followed during the measurement process. Sometimes the output from a cytometer is opaquely processed and transformed by different software packages, and unless these modification can be reverted, it is difficult to use these models in such situations.

We would also highlight that there is a need for bead sets tagged at much lower MESF values. In Figure \ref{fig:gain90_mean_sd} for instance, such beads can help characterize the curve much better in the low MESF regime.

\subsection*{Acknowledgments}

This work is a contribution of the National Institute of Standards and Technology and is not subject to copyright in the United States. P.B. was funded through the NIST PREP grant 70NANB18H162. The authors wish to thank Bradley Alpert, Arvind Balijepalli, and Anne Talkington for useful discussions during preparation of this manuscript.

\subsection*{Research Involving Human Participants and/or Animals}

Not applicable

\subsection*{Data Availability}

Analysis scripts and data developed as a part of this work are available upon
reasonable request.

\bibliography{main}

\begin{thebibliography}{10}
\providecommand \doibase [0]{http://dx.doi.org/}%

\bibitem{mizrahi2018quantitative}
Mizrahi O, Ish~Shalom E, Baniyash M, Klieger Y. Quantitative flow cytometry: concerns and recommendations in clinic and research. {\it Cytometry Part B: Clinical Cytometry} 2018\string; 94(2)\string: 211--218.

\bibitem{steen1992noise}
Steen HB. Noise, sensitivity, and resolution of flow cytometers. {\it Cytometry} 1992\string; 13(8)\string: 822--830.

\bibitem{wang2017flow}
Wang L, Hoffman RA. Flow cytometer performance characterization, standardization, and control. {\it Single Cell Analysis: Contemporary Research and Clinical Applications} 2017\string: 171--199.

\bibitem{wang2017standardization}
Wang L, Hoffman RA. Standardization, calibration, and control in flow cytometry. {\it Current {P}rotocols in {C}ytometry} 2017\string; 79(1)\string: 1--3.

\bibitem{mittag2009basics}
Mittag A, T{\'a}rnok A. Basics of standardization and calibration in cytometry--a review. {\it Journal of {B}iophotonics} 2009\string; 2(8-9)\string: 470--481.

\bibitem{perfetto2012quality}
Perfetto SP, Ambrozak D, Nguyen R, Chattopadhyay PK, Roederer M. Quality assurance for polychromatic flow cytometry using a suite of calibration beads. {\it Nature {P}rotocols} 2012\string; 7(12)\string: 2067--2079.

\bibitem{finck2013normalization}
Finck R, Simonds EF, Jager A, et al. Normalization of mass cytometry data with bead standards. {\it Cytometry Part A} 2013\string; 83(5)\string: 483--494.

\bibitem{byrd2015polystyrene}
Byrd T, Carr KD, Norman JC, Huye L, Hegde M, Ahmed N. Polystyrene microspheres enable 10-color compensation for immunophenotyping of primary human leukocytes. {\it Cytometry Part A} 2015\string; 87(11)\string: 1038--1046.

\bibitem{overton1988modified}
Overton WR. Modified histogram subtraction technique for analysis of flow cytometry data. {\it Cytometry: The Journal of the International Society for Analytical Cytology} 1988\string; 9(6)\string: 619--626.

\bibitem{hulspas2009considerations}
Hulspas R, O'Gorman MR, Wood BL, Gratama JW, Sutherland DR. Considerations for the control of background fluorescence in clinical flow cytometry. {\it Cytometry part b: Clinical cytometry: The journal of the international society for analytical cytology} 2009\string; 76(6)\string: 355--364.

\bibitem{hoffman2007characterization}
Hoffman RA, Wood JC. Characterization of flow cytometer instrument sensitivity. {\it Current Protocols in Cytometry} 2007\string; 40(1)\string: 1--20.

\bibitem{chase1998resolution}
Chase ES, Hoffman RA. Resolution of dimly fluorescent particles: A practical measure of fluorescence sensitivity. {\it Cytometry} 1998\string; 33(2)\string: 267--279.

\bibitem{ross2019first}
Ross S. {\it First Course in Probability, A}.
\newblock Pearson Higher Ed .
\newblock 2019.

\bibitem{ross2009introduction}
Ross SM. {\it Introduction to Probability and Statistics for Engineers and Scientists, Student Solutions Manual}.
\newblock Academic Press .
\newblock 2009.

\bibitem{kettlitz2014sensitivity}
Kettlitz SW, Moosmann C, Valouch S, Lemmer U. Sensitivity improvement in fluorescence-based particle detection. {\it Cytometry Part A} 2014\string; 85(9)\string: 746--755.

\bibitem{patrone2024uncertainty}
Patrone PN, Kearsley AJ, Catterton M, Cooksey G. Uncertainty Quantification in Cytometry Part I: An Analytical Perspective Beyond Q and B. {\it In preparation} 2024.

\bibitem{catterton2024uncertainty}
Catterton MA, DiSalvo M, Patrone PN, , Cooksey GA. Uncertainty Quantification for Cytometry Part II: Comparison of Serial and Traditional Cytometers. {\it In preparation} 2024.

\bibitem{wang2021establishing}
Wang L, Bhardwaj R, Mostowski H, et al. Establishing CD19 B-cell reference control materials for comparable and quantitative cytometric expression analysis. {\it PLoS One} 2021\string; 16(3)\string: e0248118.

\bibitem{gaigalas2021sources}
Gaigalas AK, Zhang YZ, Tian L, Wang L. Sources of Variability in the Response of Labeled Microspheres and B Cells during the Analysis by a Flow Cytometer. {\it International Journal of Molecular Sciences} 2021\string; 22(15)\string: 8256.

\end{thebibliography}
%\nocite{*}

\section*{Appendix}
\section*{Theory}
\emph{A shorter version of this theory is a part of Section \ref{Section:Theory} in the manuscript.}

We assume that a population of calibration beads of same size, tagged volumetrically with different quantity of fluorophores are processed through a flow cytometer and corresponding areas under the curve are obtained. We term this area under the curve to be the measured signal for a given bead and denote it by $x$. Let $f$ be the true underlying number of fluorophores for that bead.

True signal accumulates uncertainties and shows up as measured signal. We explicitly describe the uncertainties associated with bead populations and the cytometer, and how they convolve together to form the observed signal. We then use these models to estimate bead and instrument parameters.

A measured signal $x$ is obtained due to the underlying true tagged number of fluorophores on the bead, autofluorescence of the bead, as well uncertainties introduced by the act of measurement itself. As these uncertainties are not known a priori, same measurement $x$ can be generated by different values of $f$. 
That is,
\begin{equation}
x = alpha f_1 + \epsilon_1 = alpha f_2 + \epsilon_2 = \cdots
\end{equation} 

As a result, to model the distribution of measurements $x$, we need to consider all the ways a given measurement value $x$ could have been obtained from different underlying true signal values, weighing it by the probability of the original number of fluorophores being $f$. That is, using the law of total probability, the probability density of the measurement value is given as
\begin{equation}
Prob(x) = \int Prob(x | f) Prob(f) df
\end{equation}
where the integral is over the real line, $\mathbb{R}$. In the following subsections, we model the probability density $Prob(f)$ corresponding to distribution of beads and the probability density $Prob(x|f)$ corresponding to measurement-induced variation due to instrument properties.

\subsection*{Bead Uncertainties}

We start with bead populations tagged with different concentrations but otherwise identically produced, beadtypes $j = \{0,1,2,\cdots m\}$. Consider a population of $n_j$ `identical' fluorescent labeled beads of type $j$, each with $c_j$ nominal average tagged fluorophores per unit volume. The manufacturing process of such beads aims at producing beads with identical radii and tagged fluorophores \footnote{In cases where mixed bead sets are used and manufacturing processes and/or materials may not be the same, the volume, tagging, autofluorescence values for each beadtype must be included separately.}. However, these physical production processes lead to variations in volume $\sigma_v$ as well as shot noise during the process of tagging, $\sigma_t$. It is reasonable to assume that $q_j$, the true number of fluorophores for a bead of type $j$, is normally distributed around some mean $\mu_{b_j}$ number of fluorophores\footnote{This discrete quantity can be modeled as the sum of binomial random variables for small subvolumes of the bead, which can then be approximated by a normal distribution, which we model here.}, with standard deviation $\sigma_{b_j}$. These values are assumed to be unknown but invariant for a given bead type. Therefore, \footnote{Note that if the beads are tagged on the surface instead of volumetrically, we need to consider surface area variation for tagging and volume variation for autofluorescence of the beads and the terms will change.}

\begin{equation}
F_j \sim N\left( \mu_{b_j}, \sigma_{b_j}^2 \right)
\end{equation}

Each bead generates a signal due to the fluorophores on that bead. Let $c_j$ be the average number of fluorophores on bead type j.

 Notice that for $i$th bead, 
\begin{equation}
f = c_j + \text{ deviation from the mean}.
\end{equation}

This deviation can arise from various factors, such as

\begin{enumerate}
\item Due to variation from mean bead volume- We assume that the volume of these beads is distributed normally around mean volume $\bar{v}$ with variance $\sigma_V^2$. If the beads are tagged volumetrically, i.e. tagged fluorophores are distributed throughout the bead volume instead of solely on the surface, then this volume variation means that the beads will have variable amounts of tagged fluorophores given by $c_j + c_j N_V(0,\sigma_V^2)$. \footnote{If the beads in question are tagged on the surface, some modifications to this assumption are needed.} Moreover, the autofluorescence of these beads will also vary due to changes in size. Let $k_a$ be the autofluroscence factor for the bead material so that $k_a \bar{v}$ is the mean autofluorescence. Then, the fluorophores due to autofluorescence is distributed as  $k_a \bar{v} +k_a N_V(0,\sigma_V^2).$ In order to fix the scale of the problem, we set the average volume of a bead to $\bar{v}=1$.

\item Due to variation introduced by tagging of the beads with fluorophores- The physical processes involved in tagging of the beads introduce a shot noise error in the number of tagged fluorophores, with variance $\sigma_t^2$, given by $\sqrt{c_j}N_t(0,\sigma_t^2)$

\end{enumerate}

Notice that the realizations of normal are distinct from each other, we assume independence between the volume and tagging processes. Using both these variations, the true number of fluorophores on a given bead is the mean number of fluorophores on a bead due to tagging and autofluoroscence with a variance in terms of the volume and tagging variances,  
\begin{equation}
F_j \sim N\left( c_j + k_a, (c_j + k_a)^2 \sigma_V^2 + c_j\sigma_t^2\right).
\end{equation}

\subsection*{Instrument Uncertainties}

The distribution of a measurement   $X_j$ given an underlying true number of fluorophores $F_j$ is modeled next. Notice that in an ideal instrument, the instrument will measure the exact underlying signal, that is, the distribution of $X_j$ given $F_j=f_j$ will just be a Dirac-delta peak at a single point. This is of course not physically feasible, and the act of measurement itself changes the distribution of observed signal. The true signal is translated due to background $B$ before measurement. Moreover, the photodetector endows shot noise $\sigma_s$ onto the measurement that depends on the true underlying fluorophores $f_j$, background $B$, gain factor $G$, gain-independent variance $\omega$. This helps us model the measurement given a bead true intensity is distributed as
\begin{equation}
X_j|(Q_j=f_j) \sim N(G(f_j+B), G^2\sigma_s^2( f_j+B) + \omega)
\end{equation}

This assumes a square dependence of variance on gain factor. 

Moreover, there is a observed and well-known phenomenon of non-gain dependent thermal noise variance that is accounted through the factor $\omega$.

\subsection*{Model for measurements}

Altogether, using eqs \eqref{eq: Model Total Prob},\eqref{eq: Model Q}, and \eqref{eq: Model X given Q} for the underlying models, we obtain the probability density function for the distribution of measurements as follows: 
Given the vector of parameters $\bm{\theta}$, the probability density function $g_{\bm{\theta}}$ is given below. 

\begin{equation}
g_{\bm{\theta},j}(x) = \frac{1}{2\pi\sigma_{b_j} } \int \frac{1}{\sqrt{G^2\sigma_s^2( f_j+B) + \omega}}\ e^{-\frac{1}{2} \left(\frac{(x_{i,j}-Gf_j-GB)^2}{G^2\sigma_s^2( f_j+B) + \omega} + \frac{(f_j-\mu_{b_j})^2}{\sigma_{b_j}^2}\right)} df_j,
\label{eq:fullmodel}
\end{equation}
where $x_{i,j}$ is the MFI for $i^{th}$ bead of type $j$. All integrals are over the real line. \emph{For the ease of reading, we drop the indexing by $j$ for the rest of this subsection.}

Integrating this function over a range of values $x$ gives you the probability of the actual measurement belonging to that set. On the other hand, this function is also used to write down a likelihood function so that maximizing this function for the given dataset gives us a vector of parameters that are the most likely. Also note that this definition of $g$ involves an integral that we haven't found an analytic expression for after considerable efforts. Note that the sticking points are the $f$ dependence in the denominator of the function, as well as of the exponent. In fact, even numerical integration is rather time-consuming as well as error prone. As a result, we approximate the integral under some reasonable assumptions. These assumptions and the derivations are detailed here.

We perform a change of variables, with $x = \tilde{x}G$, and $\tilde{\omega} = \frac{\omega}{G^2}$ to obtain
\begin{equation}
g_{\bm{\theta}}(x) = \frac{1}{2\pi\sigma_{b}} \int \frac{1}{\sqrt{G^2\sigma_s^2( f+B) + \omega}}\ e^{-\frac{1}{2} \left(\frac{(\tilde{x}-f-\tilde{B})^2}{\sigma_s^2( f+B) + \tilde{\omega}} + \frac{(f-\mu_b)^2}{\sigma_b^2}\right)} df 
\end{equation}
  
Under the assumptions that the beads have a low CV or variance i.e. by assuming that the beads are all reasonably manufactured and tagged, we can approximate the distribution of $X|(F=f)$ by assuming that the variance of this distribution can be approximated by the variance at $\mu_b$. We obtain

\begin{equation}
\tilde{g}_{\bm{\theta}}(x) = \frac{1}{2\pi\sigma_{b}\sqrt{G^2\sigma_s^2( \mu_b+B) + \omega}} \int\ e^{-\frac{1}{2} \left(\frac{(\tilde{x}-f-\tilde{B})^2}{\sigma_s^2( \mu_b+B) + \tilde{\omega}} + \frac{(f-\mu_b)^2}{\sigma_b^2}\right)} dq. 
\label{eq: approx1}
\end{equation}

For ease of notation, let
\begin{equation}
C_1 = \frac{1}{\sigma_s^2( \mu_b+B) + \tilde{\omega}}, \ C_2 = \frac{1}{\sigma_b^2}.
\label{eq: C1C2}
\end{equation}

We focus on the exponent now. We add and subtract $\mu_b$ in the first term in the exponent and simplify using completing the square to obtain 

\begin{align*}
C_1 (\tilde{x}-f-B)^2 + C_2(f-\mu_b)^2 & = C_1 ((\tilde{x}-B-\mu_b)-(f-\mu_b))^2 + C_2(f-\mu_b)^2\\
& = C_1 (\tilde{x}-B-\mu_b)^2 - 2 C_1 (\tilde{x}-B-\mu_b)(f-\mu_b) + (C_1+C_2) (f-\mu_b)^2\\
& = \left(\sqrt{C_1+C_2} (f-\mu_b)-\frac{C_1}{\sqrt{C_1+C_2}}(\tilde{x}-B-\mu_b)\right)^2 \\
& \ \ \ + \left(C_1 - \frac{C_1^2}{C_1+C_2}\right) (\tilde{x}-B-\mu_b)^2\\
& = (C_1+C_2)\left((f-\mu_b)-\frac{C_1}{C_1+C_2}(\tilde{x}-B-\mu_b)\right)^2 \\
& \ \ \ + \left(C_1 - \frac{C_1^2}{C_1+C_2}\right) (\tilde{x}-B-\mu_b)^2.
\end{align*}
Now, we plug this simplification back into equation \eqref{eq: approx1}. Using the properties of Gaussian integral and the identity $\int e^{-a(y+b)^2} dy = \sqrt{\frac{\pi}{a}}$, this calculation follows,

\begin{align*}
\tilde{g}_{\bm{\theta}}(x) & = \frac{e^{-\frac{1}{2}\left(C_1 - \frac{C_1^2}{C_1+C_2}\right) (\tilde{x}-B-\mu_b)^2}}{2\pi\sigma_{b} \sqrt{G^2\sigma_s^2( \mu_b+B) + \omega}} \int\ e^{-\frac{C_1+C_2}{2} \left((f-\mu_b)-\frac{C_1}{C_1+C_2}(\tilde{x}-B-\mu_b) \right)^2} df\\
& = \frac{e^{-\frac{1}{2}\left( \frac{C_1C_2}{C_1+C_2}\right) (\tilde{x}-B-\mu_b)^2}}{2\pi\sigma_{b} \sqrt{G^2\sigma_s^2( \mu_b+B) + \omega}} \sqrt{\frac{2\pi}{C_1 + C_2}}\\
& = \frac{1}{\sqrt{2\pi} \sqrt{G^2\sigma_b^2 + G^2 \sigma_s^2(\mu_b+B)+\omega}} e^{-\frac{1}{2}\frac{(x-GB-G\mu_b)^2}{G^2\sigma_b^2 + G^2 \sigma_s^2(\mu_b+B)+\omega} }\end{align*}

where the last step uses the definitions of $C_1, C_2, \tilde{x}, B$ to obtain
\begin{equation}
\frac{1}{C_1+C_2} = \frac{\sigma_b^2(\sigma_s^2(\mu_b+B)+\tilde{\omega})}{\sigma_b^2 + \sigma_s^2(\mu_b+B)+\tilde{\omega}} \text{ and } \frac{C_1 C_2}{C_1+C_2} = \frac{1}{\sigma_b^2 + \sigma_s^2(\mu_b+B)+\tilde{\omega}}
\end{equation}
 Notice that the simplified form of $\tilde{g}_{\bm{\theta}}$ is a normal pdf. Thus, under the assumptions described earlier, the measurements for beadtype $j$ are distributed as 
\begin{equation}
X_j \sim N(G(B+\mu_{b_j}),G^2 \sigma_{b_j}^2 + G^2 \sigma_s^2 (\mu_{b_j}+B)+\omega).
\end{equation}

This approximation has a much simpler distribution, easier to calculate while still including the crucial fact that the uncertainty of measurement inherently depends on the magnitude of the underlying true signal. From this point onward, we utilize this approximation. Care must be taken and the full model used while considering the cases where the low CV assumption may not hold true.

\subsection*{Estimation of Parameters}
Let $\bm{\theta} = [B\  \sigma_s^2\  c\  k_a\  \sigma_V^2\  \sigma_t^2]$. \footnote{Notice that we haven't included the volume of the beads as a factor here. We assign $\bar{v} = 1$ as essentially a way to fix the scale of the problem. This is important to remember, especially if running the optimization with beads of different sizes, that factor needs to be accounted for.} The probability of a particular measurement depends on these parameters. Let $\mathcal X = \{\mathcal X_j\}$ denote the set of measurements, where $\mathcal{X}_j=\{x_{i,j}:i=1,\cdots,n_j\}$ for beads of type $j\in \mathcal{J} = \{0,1,2,\cdots,m\}$, and denote the set of measured gains by $\mathcal{G}$, the negative log likelihood function is
\begin{align*}
\mathcal{L}(\bm{\theta}|\mathcal{X}) & = -\ln \prod_{G \in \mathcal{G},\ j\in \mathcal{J},\ i=1}^n \tilde{g}_{\bm{\theta}}(x_i)\\
& = -\sum_{G \in \mathcal{G},\ j\in \mathcal{J},\ i=1}^n \ln \tilde{g}_{\bm{\theta}}(x_i)\\
& = \sum_{G\in \mathcal{G}, j\in \mathcal{J}} \left( \frac{n_j}{2} \ln(2\pi)+ \frac{n_j}{2} \ln(G^2 \sigma_{b_j}^2 + G^2 \sigma_s^2 (\mu_{b_j}+B)+\omega) + \frac{1}{2} \sum\limits_{i=1}^n \frac{(x_{i,j}-G(B+\mu_{b_j}))^2}{G^2 \sigma_{b_j}^2 + G^2 \sigma_s^2 (\mu_{b_j}+B)+\omega}\right).
\end{align*}

We minimize this simultaneously for all beads to obtain an estimate of parameters. We regularize this by giving higher weight to lower beadtypes so as to get better estimates of quantities such as autofluorescence and background, as signal dominates over noise in such a regime. More details about the level sets of the objective function and their resolution in the next section.

\subsubsection*{Scaling to identify level sets}

\emph{For the ease of reading, we drop the indexing by $j$ for the rest of this subsection.}

Notation: $\mu = G(\mu_b+B) = G(B+k_a+c)$.

We quickly realize that this objective function is highly non-convex over the parameter space and therefore it is important to identify the manifold over which the minimum is attained. We first scale the model with respect to $\sigma_v^2$ using
\begin{equation}
\hat{c} = \sigma_V c,\ \hat{k_a} = \sigma_V k_a,\ \hat{B} = \sigma_V B,\ \hat{\sigma}_t^2 = \frac{\sigma_t^2}{\sigma_V},\ \hat{\sigma}_s^2 = \frac{\sigma_s^2}{\sigma_V},\ \hat{\mu} = \sigma_V \mu.
\end{equation}
to obtain

\begin{equation}
\hat{\mu} = G(\hat{B} + \hat{k_a} + \hat{c}) \quad \text{ and } \quad \sigma_m^2 = G^2(\hat{c}+\hat{k_a})^2 + G^2 \hat{c} \hat{\sigma}_t^2 + G \hat{\sigma}_s^2 \hat{\mu} + \omega.
\end{equation}

 In order to identify the level sets, we first assume scaling of every variable as follows:
 \begin{equation}
 \hat{B} \rightarrow \alpha \hat{B},\ \hat{k_a} \rightarrow y \hat{k_a},\ \hat{c}\rightarrow x \hat{c},\ \hat{\sigma_t}^2 \rightarrow z \hat{\sigma_t}^2,\ \hat{\sigma_s}^2 \rightarrow S \hat{\sigma_s}^2, \omega \rightarrow W\omega. 
 \end{equation}
 
 We know that if the means and variances for each bead match at each gain for two different parameter values then both those vectors of parameter values will fall on the same level set. To that end, we set up the following:
 
\begin{equation}
\hat{\mu} = G(\hat{B}+\hat{k}_a+\hat{c}) = G(\alpha \hat{B}+y \hat{k}_a+ x \hat{c}).
\end{equation}

We can also write
\begin{equation}
\sigma_m^2 = G^2(\hat{c}+\hat{k}_a)^2 + G^2 \hat{c} \hat{\sigma}_t^2 + G \hat{\sigma}_s^2 \hat{\mu} + \omega = G^2(x \hat{c}+y \hat{k_a})^2 + G^2 x \hat{c} z \hat{\sigma}_t^2 + G S \hat{\sigma}_s^2 \hat{\mu} + W\omega.
\end{equation}
In other words, for any gain $G$,
\begin{equation}
G^2\left((x \hat{c}+y \hat{k_a})^2 - (\hat{c}+\hat{k}_a)^2\right) + G^2  \hat{c} \hat{\sigma}_t^2 \left(x z - 1 \right) + G (S-1) \hat{\sigma}_s^2 \hat{\mu} + (W-1)\omega = 0.
\end{equation}
As this holds for any gain $G$, we can thus conclude that 
\begin{equation}
(S-1) \hat{\sigma}_s^2 \hat{\mu} = 0 \Rightarrow S = 1, (W-1)\omega = 0 \Rightarrow W = 1
\end{equation}
and also
\begin{equation}
\left((x \hat{c}+y \hat{k_a})^2 - (\hat{c}+\hat{k_a})^2\right) +  \hat{c} \hat{\sigma}_t^2 \left(xz - 1 \right) = 0.
\end{equation}

This gives us a manifold in terms of $x, y,$ and $z$,

\begin{equation}
0 = (x \hat{c}+y \hat{k}_a)^2 + \hat{c} \hat{\sigma}_t^2 x z - (\hat{c}+\hat{k}_a)^2 - \hat{c} \hat{\sigma}_t^2 = \hat{c}^2 x^2 + 2\hat{c}\hat{k}_a xy + \hat{k}_a^2y^2 + c \hat{\sigma}_t^2 xz - (\hat{c}+\hat{k}_a)^2 - \hat{c} \hat{\sigma}_t^2.
\end{equation}

Also notice the other manifold, a plane in terms of $x,y,\alpha$,
\begin{equation}
(\alpha-1)\hat{B} + (y-1) \hat{k}_a + (x-1) \hat{c} = 0 \Rightarrow  \hat{c} x + \hat{k}_a y + \hat{B} \alpha = \hat{c} + \hat{k}_a +\hat{B}.
\end{equation}

As a result, we conclude that autofluorescence, tagged intensity, tagging and volume variance are intrinsically linked and cannot be  separated from each other without additional measurements. Note that $S=1$ implies that provided there is enough data for different gains, this model can separate the instrument uncertainty from the rest and $W=1$ implies that the same is true for instrument variance factor independent of gain. Notice that there is an implicit restriction on $B$ due to the equation for $\mu$. This means that knowing any two out of $k_a, c, \sigma_t^2$ will help determine a unique minimum.

Remember that we have multiple bead-types of same size and tagged in the same way. If one of the bead-types is the untagged bead where $c=0$ and thus $\hat{c}=0$, for this bead type, the manifolds reduce down to 
\begin{equation}
0 = \hat{k}_a^2y^2 - \hat{k}_a^2 \Rightarrow y = 1 \text{ and } (\alpha-1)\hat{B} + (y-1) \hat{k}_a = 0 \Rightarrow \alpha = 1,
\end{equation}
leading to determination of $\hat{k_a}, \hat{B}$. The linear manifold then dictates the determination of $\hat{c}$ for tagged beadtypes and subsequently $\sigma_t, \sigma_v$. 
 
In order to ensure that this untagged bead distribution contributes significantly to the negative log likelihood function, regularization to the mean of untagged bead is indicated while implementing the optimization.

\textbf{More about the manifold:}
This manifold can be rewritten as
\begin{equation}
\mathbf{x}^T \mathfrak{m} \mathbf{x} = (\hat{c}+\hat{k_a})^2 + \hat{c} \hat{\sigma_t^2}, \text{ where } \mathfrak{m} =  \begin{bmatrix}
\hat{c}^2 & \hat{c}\hat{k_a} & 0.5 \hat{c} \hat{\sigma_t}^2\\ \hat{c}\hat{k_a} & \hat{k_a}^2 & 0\\ 0.5 \hat{c} \hat{\sigma_t}^2 & 0 &0
\end{bmatrix}.
\end{equation} 

To study the properties of this manifold, let
\begin{equation}
\mathfrak{M} = \begin{bmatrix}
\mathfrak{m} & \mathbf{0}\\
\mathbf{0}^T &  - (\hat{c}+\hat{k_a})^2 - \hat{c} \hat{\sigma_t}^2
\end{bmatrix}.
\end{equation}

Then, the determinant of $\mathfrak{M}$ is $0.25 \hat{c}^2 \hat{\sigma_t}^4 \hat{k_a}^2 \left((\hat{c}+\hat{k_a})^2 + \hat{c} \hat{\sigma_t}^2\right)>0.$ Also note that $\mathfrak{m}$ and $\mathfrak{M}$ are both full rank matrices in general. Eigenvalues can be found computationally, and the roots of that cubic are of different signs. Therefore this is a hyperboloid of one sheet. If $\mathcal{V}$ is the matrix of eigenvectors of $\mathfrak{m}$ and if $\mathcal{D}$ is the corresponding diagonal matrix of eigenvalues, the manifold can be rewritten as
\begin{equation}
(\hat{c}+\hat{k_a})^2 + \hat{c} \hat{\sigma_t}^2 = \mathbf{x}^T \mathfrak{m} \mathbf{x} = \mathbf{x}^T \mathcal{V}^T \mathcal{D} \mathcal{V} \mathbf{x} = \mathbf{y}^T \mathcal{D} \mathbf{y}
\end{equation}

However, in realistic parameter values space, sometimes they can numerically have lower rank due to entries being very close to zero, and then this surface is closer to two parallel planes.

To see what this manifold looks like, for synthetic data, with $\theta_{synth}$ such that $B = \sigma_s^2 = k_a = 0.1, \sigma_t^2 = 0.5, c_0 = 0, c_1 = 0.1, c_2 = 0.3, c_3 = 0.9, c_4 = 2.7$. For this, the eigenvalus of $\mathfrak{m}$ are $-0.0218,0.0086, 0.0332$, meaning this is a hyperboloid with one sheet. Also can be seen from the implicit plot below.

\begin{figure}[!htbp]
\centering
\includegraphics[width=0.8\textwidth]{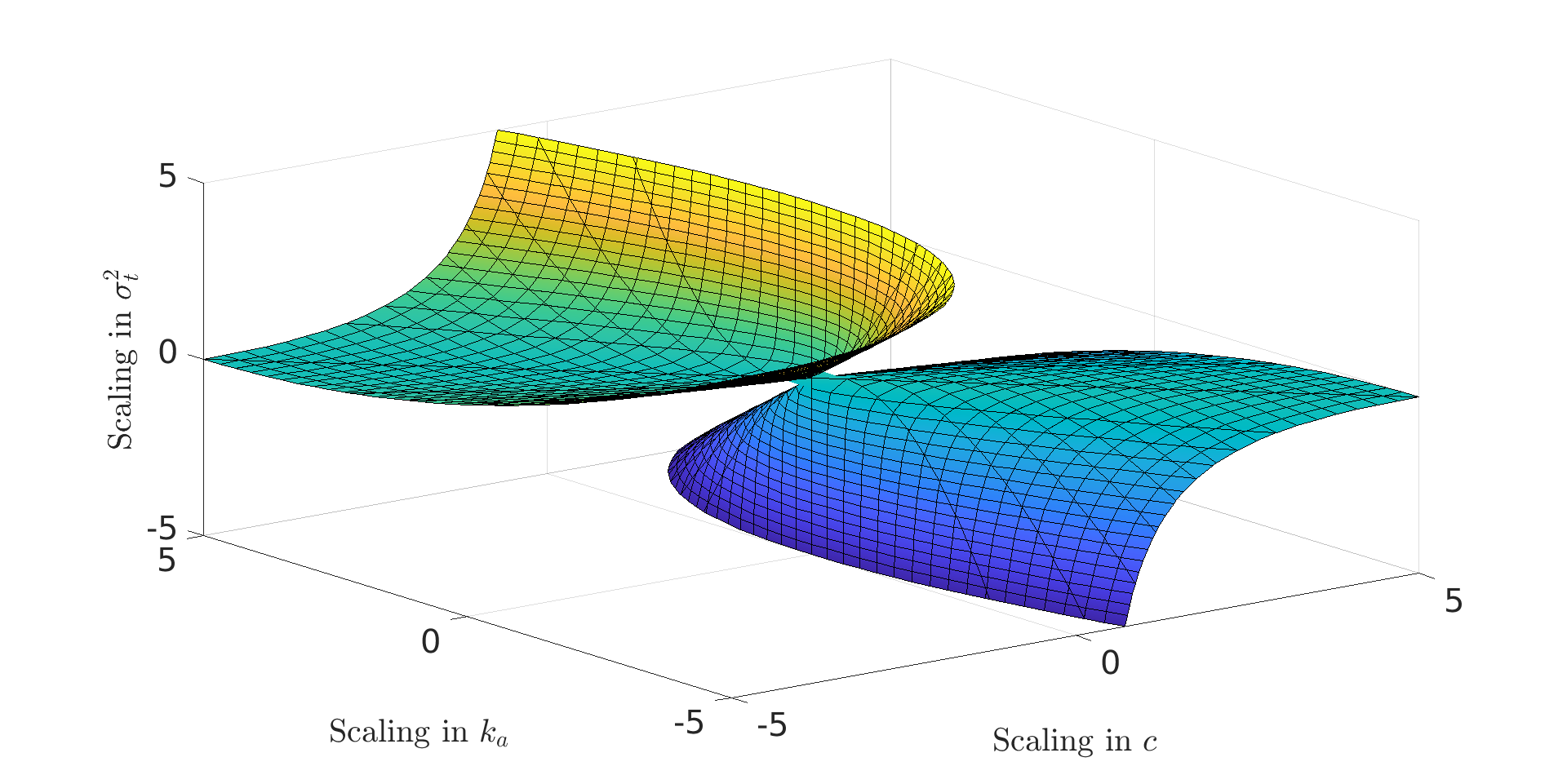}
\caption{Implicit manifold for synthetic theta. }

\end{figure}

\begin{figure}[!htbp]
\centering
\includegraphics[width=0.8\textwidth]{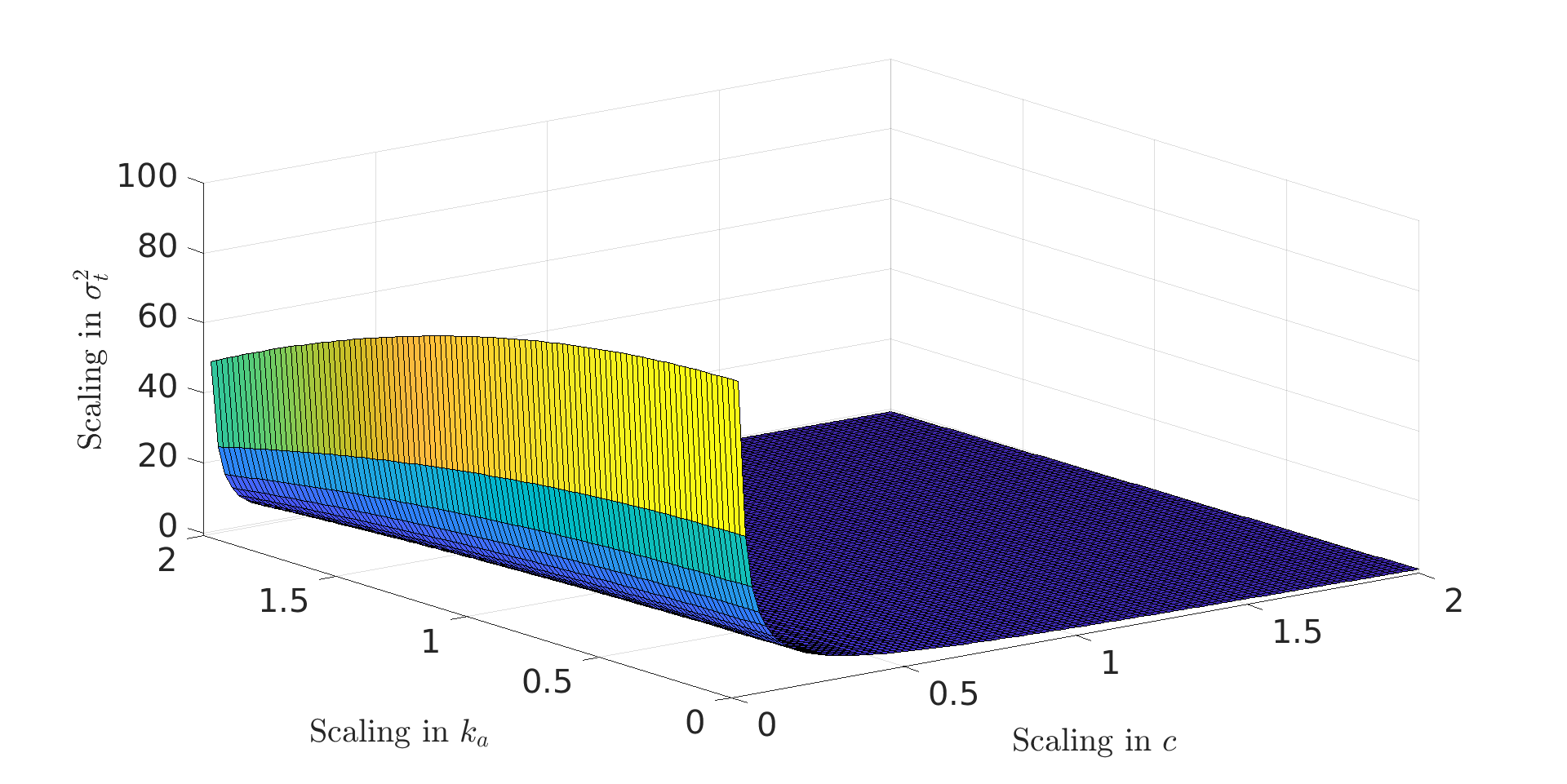}
\caption{Explicit manifold (z in terms of x,y) for synthetic theta} 
\end{figure}

\end{document}